\documentclass[useAMS,usenatbib]{mn2e}
\usepackage{amsmath}
\usepackage{amssymb}
\usepackage{graphicx}
\usepackage{longtable}
\usepackage{rotating}

\title[X-ray Detections of SMGs]{X-ray Detections of Sub-millimetre
  Galaxies: Active Galactic Nuclei Versus Starburst Contribution} 

\author[S.P. Johnson et al.]{
\parbox[t]{\textwidth}{\vspace{-1cm}S.P.~Johnson$^1$, G.W.~Wilson$^1$,
  Q.D.~Wang$^1$, C.C.~Williams$^1$, K.S.~Scott$^2$, M.S.~Yun$^1$,
  A.~Pope$^1$, J.~Lowenthal$^3$, I.~Aretxaga$^{4}$, D.~Hughes$^4$, 
  M.J.~Kim$^5$, S.~Kim$^5$, Y.~Tamura$^6$, K.~Kohno$^{6,7}$, H.~Ezawa$^8$,
  R.~Kawabe$^9$, T.~Oshima$^9$}\\ 
$^1$Department of Astronomy, University of Massachusetts, Amherst, MA
  01003, USA\\
$^2$North American ALMA Science Center, National Radio Astronomy
Observatory, Charlottesville, VA 22903, USA\\
$^3$Department of Astronomy, Smith College, Northampton, MA 01063,
USA\\
$^4$Instituto Nacional de Astrofisica, Optica y Electronica (INAOE),
Aptdo. Postal 51 y 216, 72000 Puebla, Pue., Mexico\\
$^5$Department of Astronomy \& Space Science, Sejong University,
  Seoul, Korea\\
$^6$Institute of Astronomy, University of Tokyo, Osawa, Mitaka, Tokyo
  181-0015, Japan\\
$^7$Research Center for the Early Universe, University of Tokyo,
  Hongo, Bunkyo, Tokyo 113-0033, Japan\\
$^8$ALMA Project Office, National Astronomical Observatory of Japan,
  Osawa, Mitaka, Tokyo 181-8588, Japan\\
$^9$Nobeyama Radio Observatory, National Astronomical Observatory of
  Japan, Minamimaki, Minamisaku, Nagano 384-1305, Japan}
\begin{document}
\maketitle

\begin{abstract}
We present a large-scale study of the X-ray properties and
near-IR-to-radio spectral energy distributions (SEDs) of submillimetre
galaxies (SMGs) detected at 1.1mm with the AzTEC instrument across a
$\sim$1.2 square degree area of the sky.  Combining deep $2-4$ Ms
\textit{Chandra} data with \textit{Spitzer} IRAC/MIPS and VLA data
within the GOODS-N, GOODS-S and COSMOS fields, we find evidence for
AGN activity in $\sim$14 percent of 271 AzTEC SMGs, $\sim$28 percent
considering only the two GOODS fields.  Through X-ray spectral
modeling and multi-wavelength SED fitting using Monte Carlo Markov
Chain techniques to \cite{siebenmorgen04} (AGN) and
\cite{efstathiou00} (starburst) templates, we find that while star
formation dominates the IR emission, with star formation rates (SFRs)
$\sim100-1000\rm~M_{\odot}~yr^{-1}$, the X-ray emission for most
sources is almost exclusively from obscured AGNs, with column
densities in excess of $10^{23}\rm~cm^{-2}$.  Only for $\sim$6 percent
of our sources do we find an X-ray-derived SFR consistent with 
NIR-to-radio SED derived SFRs.  Inclusion of the X-ray luminosities
as a prior to the NIR-to-radio SED effectively sets the AGN luminosity
and SFR, preventing significant contribution from the AGN template.
Our SED modeling further shows that the AGN and starburst templates
typically lack the required 1.1 mm emission necessary to match
observations, arguing for an extended, cool dust component. The cross
correlation function between the full samples of X-ray sources and
SMGs in these fields does not indicate a strong correlation between
the two populations at large scales, suggesting that SMGs and
AGNs do not necessarily trace the same underlying large scale
structure. Combined with the remaining X-ray-dim SMGs, this suggests
that sub-mm bright sources may evolve along multiple tracks, with
X-ray-detected SMGs representing transitionary objects between periods
of high star formation and AGN activity while X-ray-faint SMGs
represent a brief starburst phase of more normal galaxies.
\end{abstract}

\section{Introduction}

Large blank-field surveys made at (sub-)millimetre wavelengths have
identified a large population of bright, high-redshift galaxies
\citep[e.g.][and references
  therein]{hughes98,coppin06,bertoldi07,perera08,weiss09,scott10}.
These sub-millimetre galaxies (SMGs) are characterized by high
infrared (IR) luminosities, $\gtrsim$10$^{12}\rm~L_{\odot}$
\citep{blain04,chapman05}, and a redshift distribution peaking around
$z\sim$2 \citep{chapman05}.  SMGs are therefore believed to be the
high-redshift analogs to local ultra-luminous IR galaxies (ULIRGs) and
are possible progenitors of today's massive ellipticals
\citep[e.g.][]{smail04,chapman05}. However, SMGs at $z\sim2$ are more
numerous than local ULIRGS by several orders of magnitude and likely
dominate the total IR luminosity density at $z\sim$2
\citep{lefloch05,perez05,hopkins10}. The origin of these luminous,
high-redshift sources is still under debate due, in part, to the low
angular resolution at (sub-)millmetre wavelengths of current
instruments and the relative faintness of likely
counterparts. Multi-wavelength and IR spectroscopic follow-up studies
of SMGs using \textit{Spitzer} \citep[see, for
  example, ][]{men07,pope08,nardini08} suggest that SMGs are largely
dust-obscured starburst systems with star formation rates (SFRs)
$\sim$1000$\rm~M_{\odot}~yr^{-1}$.  However, it is becoming
increasingly apparent through the high X-ray detection rate of SMGs
\citep[$\sim30-50$\%, see][]{alex05a,alex05b,laird10,georgan11} and
SMG case studies \citep[i.e.,][]{tamura10} that emission from active
galactic nuclei (AGNs) may also be a crucial component to the
energetic output of SMGs. 

The likely connection between starburst and AGN activity in SMGs
is further supported by the concurrent nature of the cosmic SFR and
black hole accretion with peaks at $z\sim$2
\citep[e.g.,][]{lefloch05,merloni04}.  Simulations of SMG formation in
a merger-driven scenario also suggest that the SMG phase
precedes rapid growth of a central AGN \citep{nara10}. SMGs may
therefore represent an important phase in galaxy evolution and may
shed light on the origin of observed relations between AGN activity
and stellar mass in local galaxies \citep[i.e. the M-$\sigma$
  relation;][]{fer00,geb00,gultekin09}.  One should be cautious,
however, in extrapolating the starburst-AGN connection to the most
extreme objects \citep[i.e. radio-loud AGN,][and references
  therein]{dicken12} though such cases are a fundamentally different
population of sources.  Unfortunately, while there are a multitude of
methods for studying AGN and star formation, disentangling their
relative contributions to a galaxy's bolometric output remains
challenging.  Obtaining redshifts and other information via
optical/ultra-violet imaging and spectroscopy is exceptionally
difficult as SMGs are both distant and optically thick \citep[see
  review by ][]{blain02}.  IR spectroscopy of SMGs typically show
strong polycyclic aromatic hydrocarbon (PAH) features associated with
star-forming regions, although there are cases of power-law-like
spectra indicative of AGN
\citep{men07,coppin08,nardini08,pope08}. Arguably, the best indicator
for AGN activity is hard X-rays ($>$2 keV), which penetrate obscuring 
dust up to the Compton thick limit (neutral hydrogen column densities
of N$_{\rm{H}}\gtrsim$10$^{24}\rm~cm^{-2}$). X-ray detections are not
uniquely attributable to AGN, however, as high SFRs may produce
numerous high-mass X-ray binaries (HMXBs) that mimic the emission
of low-luminosity AGNs.  

In the past decade there have been few studies that consider X-ray
counterparts to SMGs for evidence of AGN activity, though this number
has expanded in recent years.  \cite{alex05a,alex05b} (hereafter
A05a,b) provide the earliest analysis by examining the
\textit{Chandra} counterparts to SCUBA \citep{holland99} 850$\mu$m
identified sources in the Great Observatories Origins Deep Survey
(GOODS) North field.  In their sample of 20 SMGs with radio and
spectroscopic redshift identifications taken from \cite{chapman05},
they find that $\sim$75 percent have X-ray properties  consistent with
obscured (N$_{\rm{H}}\gtrsim$10$^{23}\rm~cm^{-2}$) AGN activity.
Accounting for SMGs without spectroscopic redshifts, they suggest that
the true X-ray detection rate may be significantly lower, $\gtrsim$28
percent. However, the A05a,b sample may contain biases introduced
through the \cite{chapman05} SCUBA source catalog, which consists of
observations of known radio sources and low signal-to-noise
(S/N$<$3.5$\sigma$) sources and thus may not be representative of the
entire bright SMG population (see also, \citealt{younger07}). Further
X-ray/SMG counterpart analysis has been provided by \cite{laird10}
(hereafter LNPS10), who find a $\sim$45  percent X-ray detection rate
to radio and/or \textit{Spitzer}-identified  SCUBA sources
\citep{pope06} with a $\sim$20-29 percent AGN identification rate
based on X-ray spectral modeling. LNPS10 find that the bolometric FIR
emission is dominated by star formation in the majority of their
sources ($\sim$85 percent) after including available \textit{Spitzer}
photometry; consistent with A05a,b and other IR studies of SMGs
\citep[i.e.][]{men07,valiante07,men09}.

More recently, the studies of \cite{georgan11} (hereafter GRC11),
\cite{hill11} and \cite{bielby12} have utilized LABOCA data in the
Extended \textit{Chandra} Deep Field South (ECDFS, \citealt{weiss09})
and William Herschel Deep Field.  The analysis of GRC11 is similar to
that of LNPS10 who also find an AGN fraction of $<26\pm 9$ percent
with the mid-IR emission dominated by starburst activity, though the
fraction of starburst-powered X-ray sources is lower than estimated by
LNPS10.  The works of \cite{hill11} and \cite{bielby12} consider a
more statistical approach, utilizing the full catalogs rather than
individual sources as in A05a,b, LNPS10 and GRC11, though find a
similar SMG/X-ray detection rate ($\sim$20 percent).  They also find
that obscured AGNs preferentially have greater sub-mm emission than
unobscured AGNs; a result confirmed through ELVA observations by
\cite{heywood12}.  \cite{lutz10} find a similar relation in the ECDFS
where the X-ray luminosity and absorbing column density for bright
AGNs, $L_{2-10\rm{keV}}\gtrsim10^{43}\rm~ergs~s^{-1}$, is correlated
with the 870$\mu$m flux, implying a close connection to star
formation. This assumes, however, that the X-ray emission is purely
from the AGN while the 870$\mu$m flux is only from star
formation. Furthermore, the \cite{lutz10} study does not account for
X-ray-bright SMGs, which may potentially bias the stacking results.  

To recap, X-ray studies to-date find that the AGN fraction of SMGs is
in the range of $\sim$20-45 percent and that the bolometric IR luminosity
of SMGs is dominated by starbursts. 

In this work, we examine the identification rate and contribution of
AGNs to the emission at various wavelength regimes in AzTEC SMGs.
Our sample consists of \textit{Chandra} X-ray counterparts to AzTEC
1.1 mm sources found in the GOODS-North, GOODS-South and COSMOS fields,
providing a total \textit{Chandra} sky coverage of $\sim$1.15 square degrees
($\sim$0.12, $\sim$0.11 and $\sim$0.92 square degrees, respectively)
with more than 2600 identified X-ray sources.  This large sample size
will reduce any biases due to cosmic variance in previous
studies. Furthermore, we do not base our sample selection and
counterpart identification on prior source association, thus removing
any possible pre-identification bias. The available multi-wavelength
photometry in these fields, including \textit{Spitzer} IRAC and MIPS,
will provide additional constraints on the AGN identification rate and
contribution to the bolometric output of our sources.

We begin with a description of the AzTEC and \textit{Chandra} data and
reduction procedures.  We then detail our method for identifying X-ray
counterparts to the AzTEC sources and subsequent multi-wavelength
counterparts.  Our analysis of the X-ray-identified AzTEC sources
follows a two-pronged approach: (1) applying X-ray spectral models and
SED templates to the X-ray spectra and near-IR-to-radio SED, which
will provide the basic information concerning the contribution of AGN
and star formation in each wavelength regime; and (2) linking the
X-ray spectral fits to the near-IR-to-radio SED modeling, thus
providing greater insight into the AGN/star formation connection. Our
SED fitting differs from typical SED analyses
(e.g. \citealt{sejeant10}) in that we employ a Monte Carlo Markov
Chain (MCMC) technique. We close by comparing the implications of our
work to those of previous X-ray/SMG results in addition to the
X-ray/SMG cross-correlation relation. Additional analysis of our data,
including source stacking and IR-optical-UV fitting, will be presented
in future publications. 

Throughout this work, we assume a flat $\Lambda$CDM cosmology with
H$_0=70\rm~km~s^{-1}~Mpc^{-1}$,$\Lambda_0=0.73$ and
$\Omega_{\rm{M}}=0.27$. 

\section{Observations and Data Processing}

\subsection{AzTEC: 1.1 mm Observations}

AzTEC \citep{wilson08} is a 144-element bolometer array operating at
1.1mm and installed on the 50m Large Millimetre Telescope
\citep[LMT;][]{schloerb08}.  Prior to its installation on the LMT,
AzTEC has performed several science observations on the James
Clerk Maxwell Telescope (JCMT) and the Atacama
Sub-millimetre Telescope Experiment (ASTE), including blank fields
(namely GOODS-N, GOODS-S and COSMOS) and high redshift radio clusters.
Here, we briefly describe the AzTEC observations and 1.1 mm source
sample that will be used in our analysis.

During the JCMT 2005-2006 observing campaign, \cite{perera08} imaged a
21$\arcmin\times15\arcmin$ area of the GOODS-N region.  During the
2007 and 2008 observation seasons on ASTE, AzTEC imaged both GOODS-S
\citep{scott10} and the one square degree area of COSMOS
\citep{Aretxaga10}.  In reducing the raw time-steams for each set of
observations, an iterative technique using Principle Component
Analysis (PCA) is used to filter out the atmospheric signal that
dominates the raw observed data.  \cite{downes11} provides a
discussion on correcting the PCA transfer function and lists revised
catalogs for previously released AzTEC data.  Here, we use the revised
catalogs of Downes et al. for GOODS-N and GOODS-S; the COSMOS
catalog of \cite{Aretxaga10} follows this prescription. The final
AzTEC maps are constructed to have uniform coverage and sensitivity,
providing a 1$\sigma$ rms of $\sim$1.3 mJy in GOODS-N and
COSMOS. The GOODS-S map reaches the confusion limit of AzTEC on ASTE
for a depth of (1$\sigma$) $\sim$0.6 mJy rms. Sources are defined as
peaks in the signal map with S/N$\ge$3.5$\sigma$, resulting in a total
sample of 277 AzTEC sources (40, 48  and 189 in GOODS-N, GOODS-S and
COSMOS, respectively) where $\lesssim$20 are expected to be false
detections. Note, however, that the false detection rate is estimated
for a S/N threshold of $\sim$3.5$\sigma$ and decreases rapidly for
higher source S/N.  For the following analysis, we use the full sample
of 277 AzTEC sources, applying no additional source-selection criteria.

\subsection{\textit{Chandra} Observations}

The \textit{Chandra} X-ray Observatory provides deep observations of
the GOODS-N, GOODS-S and COSMOS fields \citep[for details on the
  observations see][respectively]{alex03,luo08,elvis09} with total
exposure times of $\sim$2Ms in each field.  More recently, an
additional $\sim$2Ms has been added to GOODS-S with 31 additional
pointings; bringing the final integrated exposure time to $\sim$4Ms
\citep{xue11}.  Due to the pointing strategy for COSMOS, effective
exposures only reach $\sim$200ks for the inner $\sim$0.5 sq. degree
\citep[see also][]{elvis09}.  As a result, the X-ray photon statistics
in COSMOS are very poor, leading to weak constraints on the X-ray
spectral properties (\S~3.1).  This is somewhat offset by its larger
area than the GOODS fields by allowing for more potential counterparts
(\S~2.3).  On the other hand, the deep 4Ms data in GOODS-S provides
the greatest improvement to the counting statistics, and thus spectral
modeling, to date; a valuable asset for potentially faint and highly
obscured AGNs.  All of the fields were imaged with the Advanced CCD
Imaging Spectrometer Imaging array (ACIS-I), which is composed of four
CCDs arranged in a 2$\times$2 grid that operate together to provide a
$\sim 17\arcmin\times 17\arcmin$ field-of-view with sub-arcsecond
resolution at the telescope aim-point, degrading with increasing
off-axis distance.   

To ensure uniformity in our analysis, all observations were re-reduced
using \textit{Chandra} Interactive Analysis of Observations
(\textsc{ciao} version 3.4) routines and custom routines developed for
working with merged X-ray data sets; using the published X-ray
catalogs of \cite{alex03}, \cite{luo08}, \cite{xue11} and
\cite{elvis09} would have required additional calibrations for
compatibility.  Event files and exposure maps constructed in the
0.5-8.0 keV energy range were made for all observations and
then merged to produce final maps for the three fields.

We use the source detection method of \cite{wang04}, with a false
detection probability threshold of 10$^{-6}$, to produce X-ray source
lists from the final images for cross-correlation with the AzTEC
sample and spectral extraction.  This detection method uses a wavelet
analysis of the input images (in this case, the final merged X-ray
images for each field) followed by a sliding-box map detection and
maximum likelihood analysis for both source centroiding and optimal
photometry.  During the source
detection, the X-ray maps are divided into different energy bands
(i.e. 0.5-8.0 keV full band, 0.5-2.0 keV soft band and 2.0-8.0 keV
hard band), resulting in a source catalog that includes all sources
found in each energy band along with their respective count rates and
positional uncertainties. The source detection
process also produces a list of source regions, which are defined as
circular regions with radius equal to twice the 90 percent energy
encircled fraction (defined according to the PSF at the source
position).

COSMOS poses a dilemma for source detection due to the
blending of PSFs from the tiling of observations.  To avoid this
issue, we perform the X-ray source detection on the individual
observations and then combine the resulting source lists into a final
catalog.  Derived parameters are re-calculated for each source using
the final COSMOS map, with extraction radii determined from the
smallest PSF corresponding to each source.  Alternatively, one could
simply average the sub-catalogs to produce the final catalog; however,
this may exclude X-ray counts present in an image where the source was
not initially detected. Certainly, this method has difficulty in
detecting the faintest sources present in COSMOS; nevertheless, this
will not significantly influence our results given the already low
depth of COSMOS compared to the two GOODS fields.

Combining the source lists from each field results in a total of 2630
X-ray sources available for our study.  Individually, there are 478,
526, and 1626 sources in GOODS-N, GOODS-S, and COSMOS, respectively.
Despite the differences in data reduction and source detection, our
source lists recover $\gtrsim$90 percent of those from the
published catalogs of \cite{alex03}, \cite{luo08}, \cite{xue11} and
\cite{elvis09}. However, we miss many faint sources from the published
catalogs due to our more stringent false detection threshold of
$10^{-6}$ versus $\sim$1-2$\times 10^{-5}$ for the other catalogs.

\subsection{Counterpart Candidates}

\subsubsection{\textit{Chandra} Counterparts}

\begin{figure}
\includegraphics[width=0.4\textwidth]{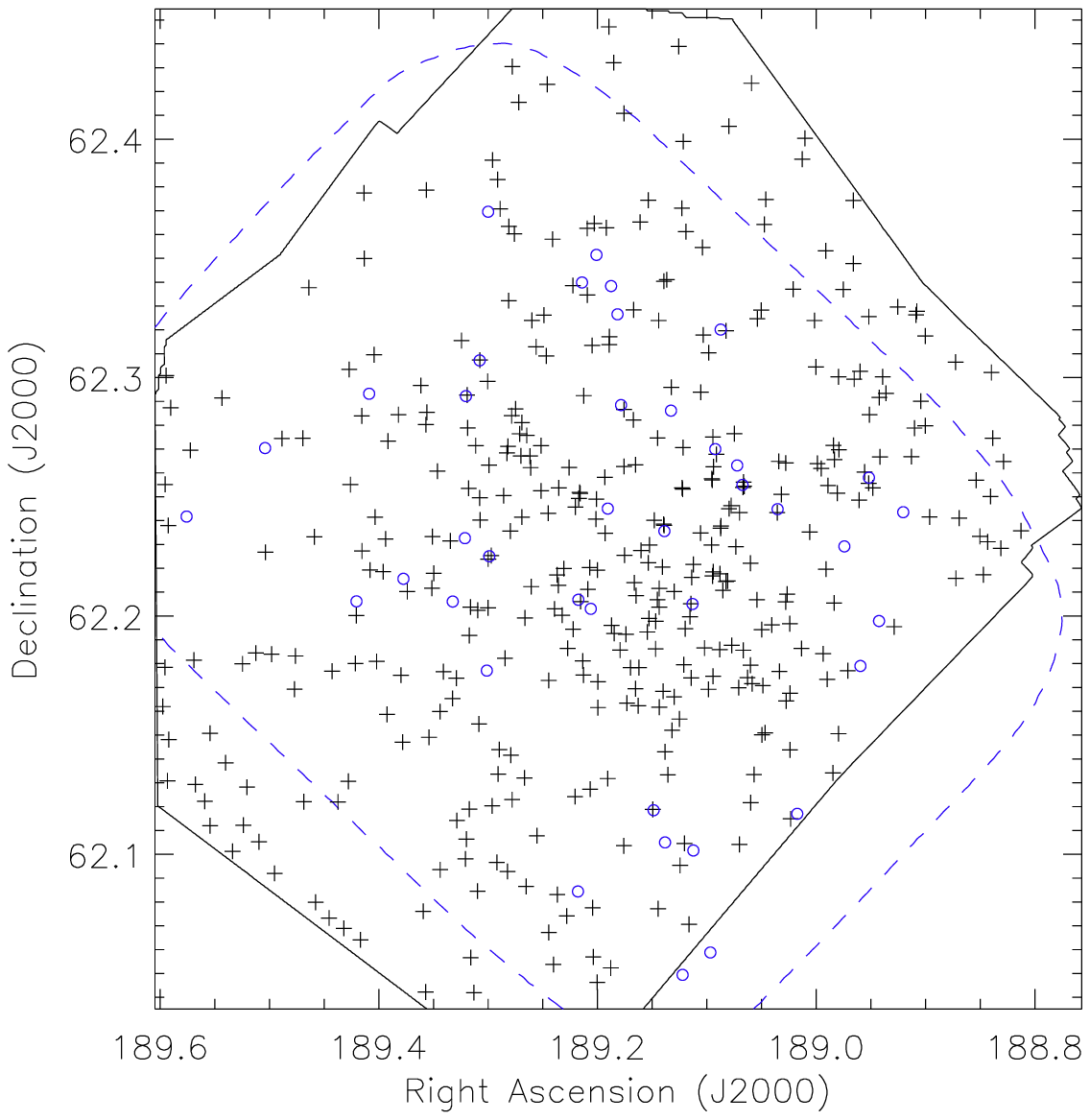}
\includegraphics[width=0.4\textwidth]{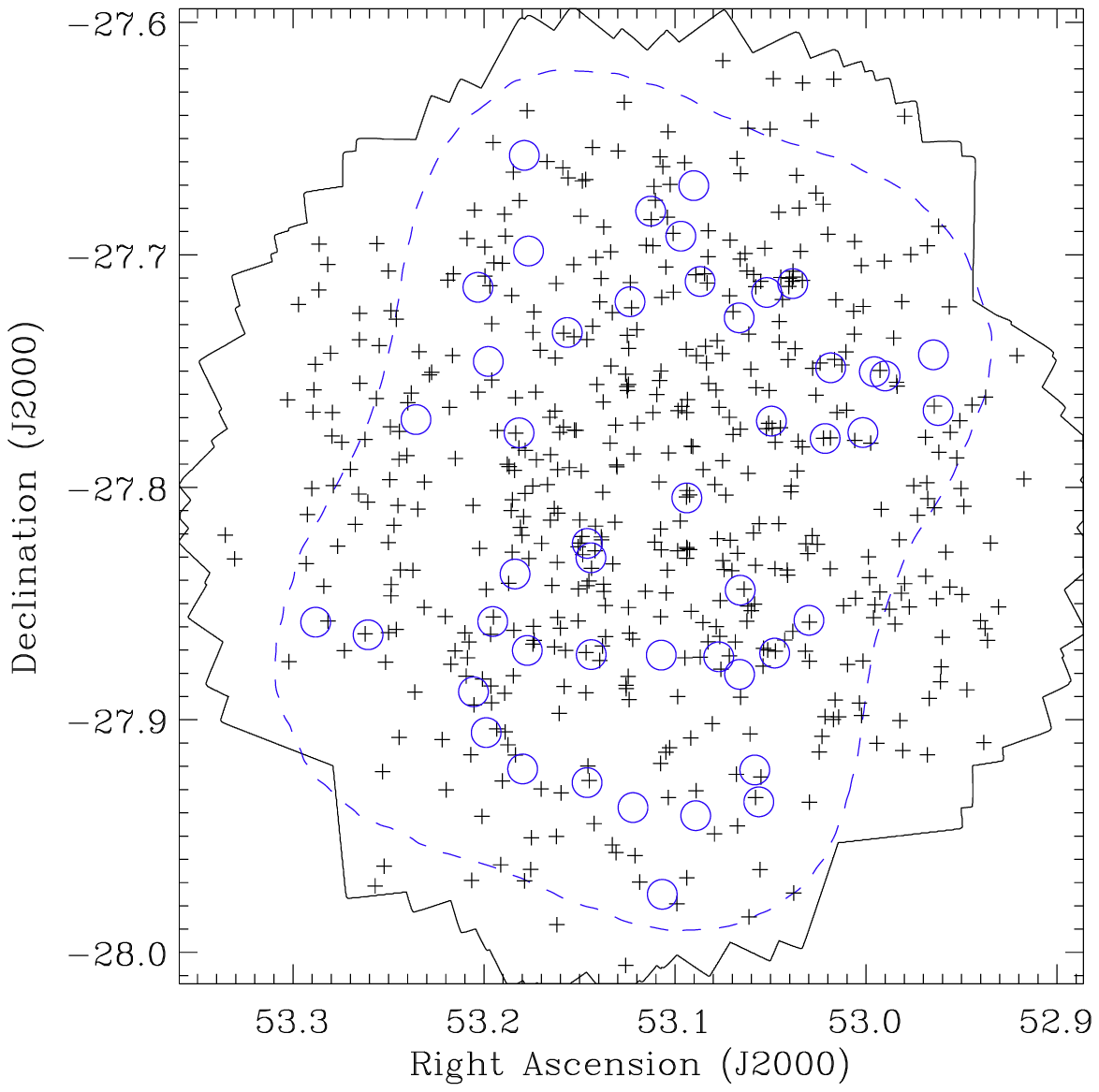}
\includegraphics[width=0.4\textwidth]{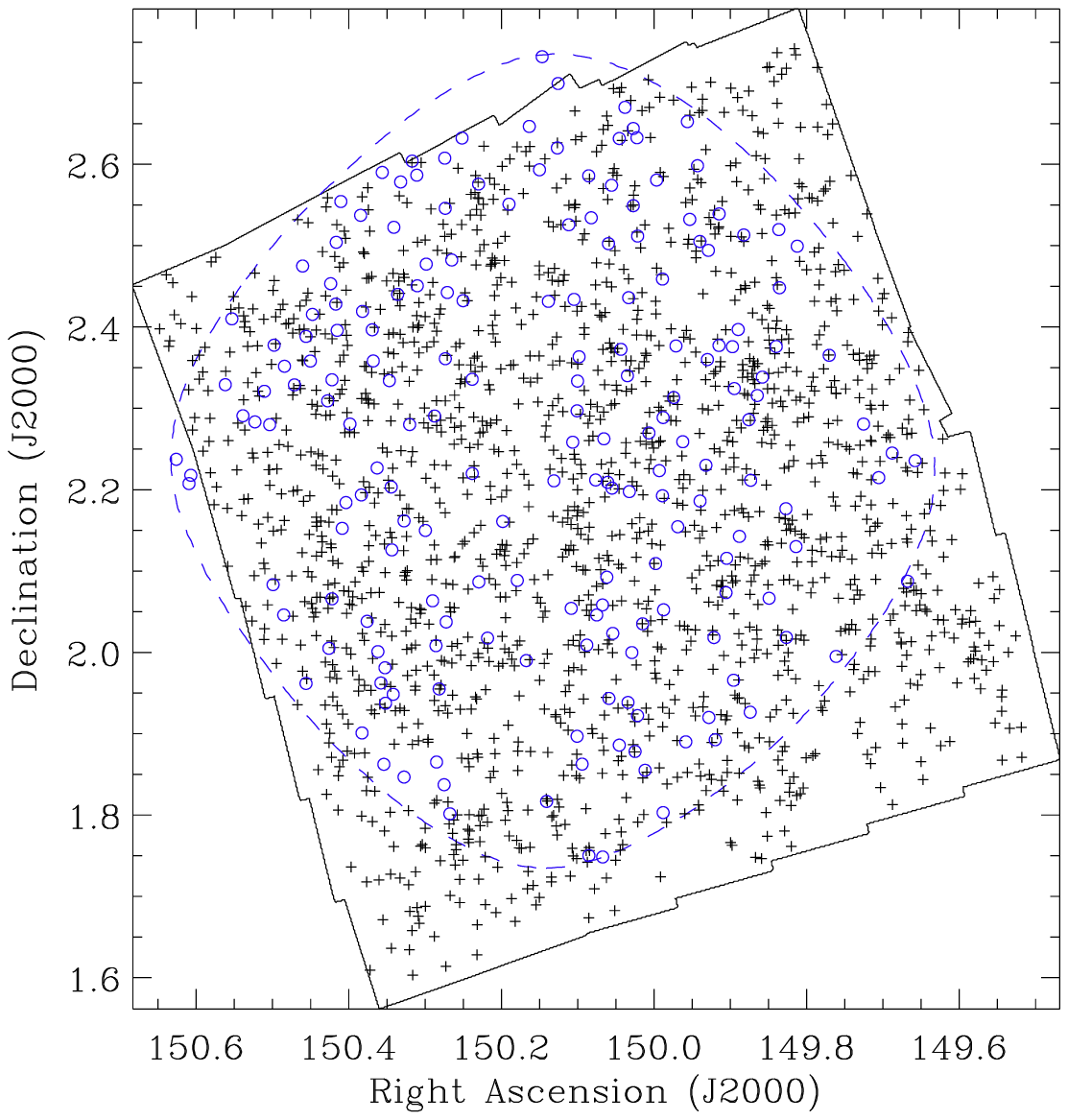}
\caption{\textit{Chandra} (solid black line) and AzTEC (dashed blue
  line) coverage regions for GOODS-N (upper), GOODS-S (middle) and
  COSMOS (lower).  The AzTEC coverage given here corresponds to the 50
  percent uniform coverage region used for source detection. Small
  circles with radii equal to the AzTEC beam-size (18$\arcsec$ in
  GOODS-N and 28$\arcsec$ in GOODS-S and COSMOS) are plotted at the
  AzTEC source positions. X-ray source positions are indicated by the
  small 'plus' symbols.}
\label{fig:overlap}
\end{figure}

The beam size of AzTEC on the JCMT and ASTE is 18$\arcsec$ and
28$\arcsec$ FWHM, respectively, making reliable X-ray counterpart
identification challenging. Following the method of \cite{chapin09}, we
use a fixed search radius of 6$\arcsec$ in GOODS-N and 10$\arcsec$ in
GOODS-S and COSMOS to find potential counterparts to the AzTEC
sources. Our choice of 10$\arcsec$ in GOODS-S and COSMOS is 
consistent with a derived search radius for a source with S/N$\sim$5.5
on the ASTE telescope according to \cite{ivison07} and roughly
corresponds to the average search radius for the AzTEC GOODS-S catalog
\citep{scott10}.  Simulations in each field agree well with
the \cite{ivison07} estimate and show that sources with
S/N$\gtrsim$3.5 are recovered within the respective search radii $>$85
percent of the time.  Extending the search radius beyond our adopted
value increases the number of X-ray counterparts; however,
these additional X-ray sources are unlikely to be true counterparts
(see below).
 
As shown in Figure~\ref{fig:overlap}, there is significant overlap
between the AzTEC and \textit{Chandra} maps.  Considering only the
overlapping regions, our sample is limited to 271 (39, 47, and 185 in
GOODS-N, GOODS-S, and COSMOS, respectively) of the initial 277 AzTEC
sources and 2229 (397, 429, 1403 respectively) of the 2630
\textit{Chandra} sources.  Of the remaining 271 AzTEC sources, we find
38 with at least one X-ray counterpart (8, 16, and 14 for GOODS-N,
GOODS-S, and COSMOS, respectively); 5 have 2 potential counterparts and
1 has 3.  For those sources with multiple potential \textit{Chandra}
counterparts, we treat each source individually and do not attempt to
split the AzTEC flux as we have no prior information on how it may be
related to the potential X-ray sources.  Overlapping spectral regions
for these sources is not an issue as the uncertainty in the X-ray
spectra is dominated by the low counting statistics.  There are a
total of 45 X-ray sources associated with the AzTEC sample, of which
only 2-3 are expected to be false identifications due to random
alignments. Comparatively, the expected number of X-ray pairs for the
entire sample of 271 AzTEC sources, assuming a purely random X-ray
source population, is $\sim$14. The AzTEC/X-ray identification rate is
therefore $\sim$14 percent, lower than estimates reported by A05a and
LNPS10 due to the shallower X-ray depth of the COSMOS field; removing
it increases the identification rate to $\sim$28 percent.   

To assess the robustness of our X-ray counterpart identifications, we
compute the probability P of random association for a given
AzTEC/X-ray pair given the search radii and X-ray source densities
(2.97, 3.14 and 1.39 $\times$10$^{-4}$ arcsec$^{-2}$ for GOODS-N,
GOODS-S and COSMOS, respectively) using the method of \cite{downes86},
which corrects for the use of a finite search radius and flux-limited
source density. The majority of the AzTEC/X-ray pairs (32/45) have
P$\le$0.05 which we define as a 'robust' counterpart, the
remaining AzTEC/X-ray pairs, with P$=$0.05-0.10, are 'tentative'
associations. Table~\ref{tab:xid} provides the list of the
\textit{Chandra}-detected AzTEC sources along with their relevant
source properties and P values.  

Through this counterpart analysis, we are implicitly assuming
that the AzTEC and X-ray source populations are physically associated
and that the two populations are not significantly clustered.  If, on
the other hand, the X-ray and SMG source populations are clustered,
then we are more likely to falsely associate sources and misinterpret
the relation between AGN and starburst systems.  \cite{almaini03}
found evidence for correlation between \textit{Chandra} and SCUBA
850$\mu$m source populations in the European Large Area ISO Survey
(ELAIS) N2 field at the 4.3$\sigma$ significance level and thus
concluded that while they trace the same large scale structure, the
AGN and starburst phases are not necessarily co-existent.  Based on
our cross-correlation analysis (see \S~4.2), we find no evidence for
significant correlation between deep \textit{Chandra} and AzTEC source
populations in general.  

\begin{table*}
  \caption{\textit{Chandra} identifications of AzTEC sources in
    GOODS-N, GOODS-S and COSMOS.  Errors are given at the 1$\sigma$
    confidence level. Col.(1): AzTEC source ID prefixed by field
    (i.e. AzGN24 for source 24 in the AzTEC GOODS-N catalog).
    Col.(2): \textit{Chandra} ID following IAU standards.  Col.(3):
    Positional offset between AzTEC and \textit{Chandra} sources.
    Errors are derived from  \textit{Chandra} positional
    uncertainty. Col.(4): \textit{Chandra} 0.5-8.0 keV full band count
    rate. Cols.(5): Total counts within the source regions as defined
    from our X-ray source detection. Col.(6): Estimated background
    counts within the source regions. Col.(7):
    deboosted AzTEC source flux (see section 3.5 of
    \citealt{austermann10} and section 6.2 of
    \citealt{scott10}). Col.(8): Probability P of the \textit{Chandra}
    source being a random association.} 
  \label{tab:xid}
  \begin{tabular}{@{}llcccccccl}
    \hline
SMM ID &\textit{Chandra} Coordinate & $\delta_x$ & 0.5-8.0 keV count
rate & Source Counts & Background Counts & 1.1mm Flux & P \\
 & (J2000) & (\arcsec) & (cnts ks$^{-1}$) & & & (mJy) \\
(1) & (2) & (3) & (4) & (5) & (6)\\
\hline
AzGN24 & J123608.57+621435.8$^\dagger$ & 5.4$\pm$0.8 & 0.031$\pm$0.007 & 98  & 61& 3.1$\pm$1.3 & 0.03\\
AzGN16a& J123615.83+621515.9$^\dagger$ & 3.1$\pm$0.5 & 0.067$\pm$0.008 & 223 & 42& 3.6$\pm$1.3 & 0.03\\
AzGN16b& J123615.93+621522.0 & 4.6$\pm$0.9 & 0.013$\pm$0.005 & 51  & 49& 3.6$\pm$1.3 & 0.02\\
AzGN16c& J123616.08+621514.1$^\dagger$ & 3.7$\pm$0.4 & 0.089$\pm$0.009 & 184 & 46& 3.6$\pm$1.3 & 0.02\\
AzGN10 & J123627.52+621218.3 & 2.7$\pm$0.5 & 0.043$\pm$0.007 & 95  & 42& 4.5$\pm$1.3 & 0.02\\
AzGN11 & J123635.86+620707.8 & 1.8$\pm$2.7 & 0.176$\pm$0.017 & 812 & 565& 4.1$\pm$1.3 & 0.01\\
AzGN14 & J123651.70+621221.7 & 4.4$\pm$0.4 & 0.222$\pm$0.015 & 301 & 42& 3.7$\pm$1.3 & 0.03\\
AzGN7a & J123711.32+621331.1$^\dagger$ & 3.3$\pm$1.0 & 0.047$\pm$0.008 & 173 & 106& 5.3$\pm$1.3 & 0.02\\
AzGN7b & J123711.98+621325.8$^\dagger$ & 4.5$\pm$1.1 & 0.043$\pm$0.008 & 150 & 100& 5.3$\pm$1.3 & 0.03\\
AzGN26 & J123713.84+621826.2$^\dagger$ & 0.5$\pm$1.5 & 0.195$\pm$0.016 & 486 & 262& 2.8$\pm$1.4 & 0.001\\
AzGN23 & J123716.63+621733.4 & 2.3$\pm$1.3 & 2.101$\pm$0.045 & 2789& 218& 3.1$\pm$1.3 & 0.01\\
AzGS29 & J033158.25-274458.8 & 9.6$\pm$2.9 & 0.079$\pm$0.013 & 1865& 1542& 2.3$\pm$0.6 & 0.09\\
AzGS8a & J033204.48-274643.3 & 8.7$\pm$1.5 & 0.201$\pm$0.012 & 1111& 650& 3.4$\pm$0.6 & 0.09\\
AzGS8b & J033205.34-274644.0 & 2.8$\pm$1.4 & 0.150$\pm$0.010 & 917 & 591& 3.4$\pm$0.6 & 0.03\\
AzGS10 & J033207.12-275128.6 & 2.9$\pm$2.2 & 0.020$\pm$0.008 & 715 & 703& 3.8$\pm$0.7 & 0.03\\
AzGS38a& J033209.26-274240.9 & 3.7$\pm$2.7 & 0.078$\pm$0.011 & 1923& 1206& 1.7$\pm$0.6 & 0.04\\
AzGS38b& J033209.71-274249.0 & 8.0$\pm$2.2 & 0.138$\pm$0.013 & 1705& 1106& 1.7$\pm$0.6 & 0.09\\
AzGS1  & J033211.39-275213.7 & 3.2$\pm$1.4 & 0.774$\pm$0.021 & 2338& 609& 6.7$\pm$0.6 & 0.03\\
AzGS13 & J033212.23-274620.9 & 5.7$\pm$0.8 & 0.247$\pm$0.012 & 789 & 260& 3.1$\pm$0.6 & 0.07\\
AzGS7  & J033213.88-275600.2 & 8.7$\pm$3.4 & 0.189$\pm$0.019 & 1932& 1497& 3.8$\pm$0.6 & 0.09\\
AzGS11 & J033215.32-275037.6 & 6.6$\pm$0.8 & 0.065$\pm$0.007 & 378 & 236& 3.3$\pm$0.6 & 0.08\\
AzGS17a& J033222.17-274811.6 & 6.6$\pm$0.3 & 0.059$\pm$0.006 & 176 & 52& 2.9$\pm$0.6 & 0.08\\
AzGS17b& J033222.56-274815.0 & 1.6$\pm$0.5 & 0.029$\pm$0.004 & 123 & 53& 2.9$\pm$0.6 & 0.01\\
AzGS34 & J033229.46-274322.0 & 9.8$\pm$1.4 & 0.027$\pm$0.006 & 492 & 392& 1.7$\pm$0.6 & 0.09\\
AzGS20 & J033234.78-275534.0 & 4.8$\pm$2.6 & 0.108$\pm$0.013 & 1853& 1490& 2.7$\pm$0.6 & 0.05\\
AzGS14 & J033235.18-275215.7 & 9.2$\pm$1.0 & 0.034$\pm$0.006 & 381 & 295& 2.9$\pm$0.6 & 0.09\\
AzGS16 & J033238.01-274401.2 & 6.3$\pm$1.6 & 0.012$\pm$0.006 & 392 & 344& 2.7$\pm$0.6 & 0.07\\
AzGS18 & J033244.02-274635.9 & 5.7$\pm$0.6 & 0.188$\pm$0.011 & 592 & 198& 3.1$\pm$0.6 & 0.07\\
AzGS25 & J033246.83-275120.9 & 6.9$\pm$1.3 & 0.041$\pm$0.007 & 521 & 400& 1.9$\pm$0.6 & 0.08\\
AzGS9  & J033302.94-275146.9 & 5.1$\pm$3.1 & 0.204$\pm$0.020 & 1421& 1097& 3.6$\pm$0.6 & 0.06\\
AzC56  & J095905.05+022156.4 & 2.7$\pm$2.6 & 0.087$\pm$0.040 & 9   & 3& 4.7$\pm$1.1 & 0.01 \\
AzC181 & J095929.70+021706.4 & 7.8$\pm$1.8 & 0.079$\pm$0.029 & 24  & 9& 2.9$\pm$1.2 & 0.04 \\
AzC101 & J095945.15+023021.1 & 6.9$\pm$3.4 & 0.284$\pm$0.065 & 56  & 29& 3.8$\pm$1.1 & 0.04 \\
AzC71  & J095953.85+021853.6 & 5.8$\pm$0.9 & 0.202$\pm$0.048 & 32  & 9& 4.3$\pm$1.1 & 0.03 \\
AzC118 & J095959.96+020633.1 & 7.0$\pm$2.3 & 0.113$\pm$0.033 & 23  & 6& 3.7$\pm$1.2 & 0.02 \\
AzC43  & J100003.73+020206.4 & 2.3$\pm$2.8 & 0.125$\pm$0.047 & 77  & 59& 4.8$\pm$1.1 & 0.009\\
AzC81  & J100006.11+015239.2 & 3.1$\pm$1.0 & 0.192$\pm$0.041 & 48  & 9& 4.1$\pm$1.1 & 0.01 \\
AzC45  & J100006.55+023259.3 & 2.2$\pm$1.4 & 0.211$\pm$0.051 & 32  & 4& 4.8$\pm$1.1 & 0.009\\
AzC44a & J100033.61+014902.0 & 3.2$\pm$0.9 & 0.303$\pm$0.054 & 55  & 5& 5.0$\pm$1.2 & 0.01 \\
AzC44b & J100033.75+014906.3 & 6.3$\pm$4.5 & 1.137$\pm$0.121 & 78  & 40& 5.0$\pm$1.2 & 0.03 \\
AzC17  & J100055.34+023441.1 & 8.6$\pm$2.1 & 4.970$\pm$0.323 & 317 & 31& 6.2$\pm$1.1 & 0.04 \\
AzC147 & J100107.46+015718.1 & 2.1$\pm$3.2 & 0.296$\pm$0.062 & 82  & 42& 3.2$\pm$1.2 & 0.007\\
AzC108 & J100116.15+023606.9 & 7.5$\pm$3.8 & 3.090$\pm$0.610 & 45  & 12& 4.0$\pm$1.2 & 0.04 \\
AzC85  & J100139.73+022548.5 & 9.0$\pm$0.8 & 0.333$\pm$0.085 & 37  & 3 & 4.0$\pm$1.1 & 0.04 \\
AzC11  & J100141.02+020404.8 & 8.7$\pm$1.8 & 0.179$\pm$0.064 & 12  & 4 & 7.9$\pm$1.1 & 0.04 \\
\hline
\end{tabular}
$^\dagger$ Source also detected in LNPS10.
\end{table*}

\begin{table*}
  \caption{VLA, \textit{Spitzer} IRAC/MIPS and redshift information
    for the X-ray identified AzTEC sources.  Spectroscopic and
    photometric redshift information for the AzTEC/X-ray sources was
    taken, primarily, from publicly available redshift catalogs (see
    \S~2.3.2 for details). MIPS upper limits are estimated from the
    5$\sigma$ upper limit of a detected MIPS source nearest the
    AzTEC/X-ray position (\S~2.3.2). Errors are given at the 1$\sigma$
    confidence level.}
  \label{tab:mid}
  \begin{tabular}{@{}llccccccrr}
    \hline
AzTEC ID & \textit{Chandra} ID & 1.4 GHz & 24 $\mu$m & 3.6 $\mu$m &
4.5 $\mu$m & 5.8 $\mu$m & 8.0 $\mu$m &$z_{spec}$ &$z_{phot}$\\
 & & ($\mu$Jy) & ($\mu$Jy) & ($\mu$Jy) & ($\mu$Jy) & ($\mu$Jy) & ($\mu$Jy) \\
\hline
AzGN24 & J123608.57+621435.8&  45$\pm$9&   51$\pm$6&   6.4$\pm$0.6&  9.5$\pm$0.8& 13.4$\pm$1.3& 18.3$\pm$1.5&      &  \\
AzGN16a& J123615.83+621515.9&  30$\pm$9&    5$\pm$7&  14.9$\pm$0.9& 19.5$\pm$0.8& 27.9$\pm$1.5& 27.1$\pm$1.7&      &  \\
AzGN16b& J123615.93+621522.0&          &           &              &             &             &             &      &  \\
AzGN16c& J123616.08+621514.1&  38$\pm$8&  326$\pm$8&  12.3$\pm$0.9& 18.1$\pm$0.8& 29.5$\pm$1.5& 43.4$\pm$1.7&  2.578&  \\
AzGN10 & J123627.52+621218.3&  18$\pm$4&   22$\pm$7&   1.2$\pm$0.4&  2.3$\pm$0.4&  4.2$\pm$1.0&  9.7$\pm$1.1&      &  \\
AzGN11 & J123635.86+620707.8& 36$\pm$10&      $<$38&   4.6$\pm$1.5&  5.6$\pm$1.5& 10.5$\pm$2.0& 22.0$\pm$2.0&   0.952&  \\
AzGN14 & J123651.70+621221.7&          &           &              &             &             &             &       &  \\
AzGN7a & J123711.32+621331.1& 127$\pm$9&  537$\pm$9&  37.9$\pm$1.2& 45.0$\pm$1.0& 53.3$\pm$1.5& 37.8$\pm$1.7&   1.996&  \\
AzGN7b & J123711.98+621325.8&  52$\pm$8&  219$\pm$7&   9.2$\pm$0.9& 11.4$\pm$0.8& 16.1$\pm$1.3& 12.3$\pm$1.5&   1.996&  \\
AzGN26 & J123713.84+621826.2& 652$\pm$5&   55$\pm$6&   3.5$\pm$0.6&  6.0$\pm$0.5&  9.4$\pm$1.3& 16.6$\pm$1.5&       &  \\
AzGN23 & J123716.63+621733.4& 381$\pm$8&1240$\pm$16&  62.7$\pm$1.2& 83.5$\pm$1.0&129.3$\pm$1.5&239.6$\pm$1.7&   1.146&  \\
AzGS29 & J033158.25-274458.8&          &      $<$80&  73.7$\pm$0.1& 49.0$\pm$0.2& 37.0$\pm$1.0& 19.9$\pm$1.0&   0.575& 0.579\\
AzGS8a & J033204.48-274643.3&          &    7$\pm$4&   3.6$\pm$0.1&  3.5$\pm$0.1&  1.3$\pm$0.6&  1.9$\pm$0.7&        & 1.450\\
AzGS8b & J033205.34-274644.0&          &  164$\pm$5&  13.4$\pm$0.1& 15.7$\pm$0.1& 20.6$\pm$0.6& 27.5$\pm$0.6&        &      \\
AzGS10 & J033207.12-275128.6&          &   26$\pm$8&   5.5$\pm$0.2&  5.7$\pm$0.2&  6.9$\pm$1.2&  4.8$\pm$1.0&        & 0.990\\
AzGS38a& J033209.26-274240.9&          &           &              &             &             &             &        &\\
AzGS38b& J033209.71-274249.0& 220$\pm$6&   39$\pm$3& 112.4$\pm$0.1& 67.6$\pm$0.1& 58.1$\pm$0.4& 34.2$\pm$0.5&   0.733& 0.762\\
AzGS1  & J033211.39-275213.7&  32$\pm$6&  122$\pm$5&  10.4$\pm$0.1& 14.6$\pm$0.1& 20.0$\pm$0.6& 28.2$\pm$0.7&       &      \\
AzGS13 & J033212.23-274620.9&          &  224$\pm$4&  53.7$\pm$0.1& 42.7$\pm$0.1& 33.1$\pm$0.4& 31.9$\pm$0.5&   1.033& 1.030\\
AzGS7  & J033213.88-275600.2&  51$\pm$6&  103$\pm$9&   7.9$\pm$0.1& 12.0$\pm$0.1& 17.7$\pm$0.6& 22.7$\pm$0.6&        &      \\
AzGS11 & J033215.32-275037.6&  46$\pm$6&  117$\pm$5&  22.9$\pm$0.1& 22.5$\pm$0.1& 23.8$\pm$0.3& 32.5$\pm$0.4&   0.250&2.280$^\dagger$\\
AzGS17a& J033222.17-274811.6&          &  200$\pm$5&  11.8$\pm$0.1& 16.5$\pm$0.1& 23.9$\pm$0.3& 20.9$\pm$0.4&        & 2.500\\
AzGS17b& J033222.56-274815.0&          &   62$\pm$7&  16.9$\pm$0.1& 20.2$\pm$0.1& 26.3$\pm$0.3& 21.2$\pm$0.4&        & 2.660\\
AzGS34 & J033229.46-274322.0&          &   70$\pm$3&  17.3$\pm$0.1& 19.9$\pm$0.1& 17.2$\pm$0.4& 14.9$\pm$0.5&        &      \\
AzGS20 & J033234.78-275534.0&          &           &              &             &             &             &   0.038&      \\
AzGS14 & J033235.18-275215.7&          &   12$\pm$3&   2.3$\pm$0.1&  3.7$\pm$0.1&  5.2$\pm$0.4& 10.0$\pm$0.4&        & 0.857\\
AzGS16 & J033238.01-274401.2&          &   46$\pm$3&   5.0$\pm$0.1&  8.1$\pm$0.1& 10.9$\pm$0.4& 16.4$\pm$0.5&   1.401& 1.180\\
AzGS18 & J033244.02-274635.9&          &  126$\pm$4&   8.2$\pm$0.1& 10.9$\pm$0.1& 16.0$\pm$0.3& 22.2$\pm$0.4&   2.688& 2.690\\
AzGS25 & J033246.83-275120.9&  90$\pm$6&  140$\pm$4&  13.9$\pm$0.1& 18.8$\pm$0.1& 24.5$\pm$0.4& 32.2$\pm$0.5&   1.101& 1.330\\
AzGS9  & J033302.94-275146.9&  87$\pm$7& 229$\pm$10&   7.7$\pm$0.1& 12.6$\pm$0.2& 14.9$\pm$0.9& 27.3$\pm$0.9&        & 3.690\\
AzC56  & J095905.05+022156.4&          &  90$\pm$10&   7.6$\pm$0.1& 11.2$\pm$0.2& 15.9$\pm$1.0& 28.0$\pm$2.5&        & 3.440\\
AzC181 & J095929.70+021706.4&          &     $<$930&  39.0$\pm$0.2& 44.9$\pm$0.3& 39.8$\pm$1.0& 26.3$\pm$2.4&        & 1.700\\
AzC101 & J095945.15+023021.1&          & 300$\pm$20&  78.1$\pm$0.2& 58.2$\pm$0.3& 44.9$\pm$1.1& 44.4$\pm$2.4&   0.893& 0.870\\
AzC71  & J095953.85+021853.6& 79$\pm$11& 520$\pm$20&  52.0$\pm$0.2& 49.2$\pm$0.3& 56.9$\pm$1.1& 44.4$\pm$2.6&   0.853& 0.720\\
AzC118 & J095959.96+020633.1&104$\pm$13& 220$\pm$20&  22.2$\pm$0.1& 23.0$\pm$0.2& 22.1$\pm$1.0& 41.5$\pm$2.1&        & 0.790\\
AzC43  & J100003.73+020206.4&          &     $<$220&   5.4$\pm$0.1&  5.6$\pm$0.2& 10.9$\pm$1.1&  8.3$\pm$2.4&        & 2.510\\
AzC81  & J100006.11+015239.2&          & 100$\pm$10&  17.5$\pm$0.1& 23.0$\pm$0.2& 19.8$\pm$0.9& 19.5$\pm$2.3&   1.796& 1.760\\
AzC45  & J100006.55+023259.3&          & 160$\pm$10&  33.9$\pm$0.2& 43.3$\pm$0.2& 51.9$\pm$1.0& 41.0$\pm$2.3&        & 1.120\\
AzC44a & J100033.61+014902.0&          & 160$\pm$20&  71.8$\pm$0.6& 61.0$\pm$0.5& 47.2$\pm$1.1& 44.8$\pm$2.2&        & 0.910\\
AzC44b & J100033.75+014906.3&          &           &              &             &             &             &        &      \\
AzC17  & J100055.34+023441.1& 78$\pm$12&1390$\pm$20&  99.7$\pm$0.2&166.1$\pm$0.4&254.9$\pm$1.2&407.6$\pm$3.0&   1.404& 1.410\\
AzC147 & J100107.46+015718.1&          &  80$\pm$10&  36.6$\pm$0.2& 35.7$\pm$0.3& 29.3$\pm$1.1& 29.3$\pm$2.3&        & 1.230\\
AzC108 & J100116.15+023606.9&          & 520$\pm$60& 128.1$\pm$0.2&140.6$\pm$0.4&162.7$\pm$1.1&188.7$\pm$2.5&   0.959& 0.950\\
AzC85  & J100139.73+022548.5&549$\pm$12& 180$\pm$20&1100.3$\pm$2.3&780.2$\pm$2.0&510.3$\pm$2.1&346.1$\pm$3.0&   0.124& 0.120\\
AzC11  & J100141.02+020404.8&          & 210$\pm$20&  11.2$\pm$0.1& 16.3$\pm$0.2& 26.5$\pm$1.0& 40.9$\pm$2.3&        &      \\
\hline
\end{tabular}

$^\dagger$ The photometric redshift was adopted for J033215.32-275037.6
  following cross-catalog comparison with GOODS-MUSIC
  \citep{santini09} and additional analysis. 
\end{table*}

\subsubsection{Multi-wavelength Counterparts}

Thanks to the extensive multi-wavelength coverage in the GOODS and
COSMOS fields, we are able to supplement the millimetre and X-ray data
of our AzTEC sample with additional photometry and
spectroscopic/photometric redshifts from the GOODS and COSMOS public
data sets.  Accurate redshifts are the most crucial given the broad
redshift distribution of SMGs and the sensitivity of X-ray spectral
modeling to redshift (\S~3.1).  Across the three fields, we utilize
publicly available VLA (1.4 GHz;
\citealt{miller08,kellermann08,morrison10}), \textit{Spitzer} IRAC
(3.6, 4.5, 5.8, 8.0 $\mu$m) and MIPS (24 $\mu$m)
SIMPLE\footnote{http://www.astro.yale.edu/dokkum/simple/},
GOODS\footnote{http://www.stsci.edu/science/goods/}, and
FIDEL\footnote{http://ssc.spitzer.caltech.edu/legacy/abs/dickinson2.html}
data, including spectroscopic/photometric redshift catalogs
where available \citep[e.g.][]{barger03,barger08,santini09,silverman10}.  
Multi-wavelength counterparts and redshifts for COSMOS were obtained
by cross-referencing our detected sources with \cite{elvis09} and the COSMOS
team's web-based data
repository.\footnote{http://irsa.ipac.caltech.edu/Missions/cosmos.html}
In cross-referencing our AzTEC/X-ray sources with other catalogs, we
use a search radius of 2$\arcsec$, the average X-ray positional
uncertainty, centered on the X-ray counterparts.  For each potential
AzTEC/X-ray pair, we find no more than one potential counterpart in
the VLA and \textit{Spitzer} catalogs; these sources have been
cross-checked with other AzTEC counterpart publications
\citep[i.e.][]{chapin09,yun12} and show excellent agreement.  For reference,
$\lesssim 1$ VLA/\textit{Spitzer} source is expected to be a
mis-association due to random alignments over all three fields. For
cases where we have IRAC but no MIPS identifications, we estimate a
5$\sigma$ MIPS flux upper limit through the photometric error of the
MIPS source nearest to the IRAC position.  A complete catalog of the
multi-wavelength photometry and redshift data for our sample is given
in Table~\ref{tab:mid}. 

\section{Analysis}

With our sample of X-ray selected AzTEC sources in hand, we now
examine their physical properties through a variety of methods.
We start with modeling of the X-ray spectra.

\subsection{X-ray Spectral Modeling}

X-ray sources with $L_{2.0-10.0{\rm{keV}}}\gtrsim
10^{42}\rm~ergs~s^{-1}$ are generally believed to be powered almost
exclusively by AGN with absorption due to modest amounts of dust and gas
within the host galaxy.  A05b showed that X-ray-identified SMGs are
predominately heavily obscured, possibly even to the Compton thick
limit with column densities of N$_{\rm{H}}\ge 10^{23}\rm~cm^{-2}$.
For the most extreme cases of obscuration, a buried AGN may only be
visible in light scattered off of the obscuring torus.  Alternatively,
if SMGs are powered by a high rate of star formation, then the
observed X-ray emission could result from the stellar population,
powered by numerous high-mass X-ray binaries (HMXB).  For comparison,
a typical SMG with SFR in the range of 100-1000
M$_{\odot}\rm~yr^{-1}$ would produce an X-ray source with
2.0-10.0 keV luminosity of $\sim10^{41-42}\rm~ergs~s^{-1}$
\citep[][hereafter P04]{persic04}. 

For our sample of AzTEC/X-ray sources, we first extract their source
and local background spectrum in the 0.5-8.0 keV observed energy range
using the region files defined from our source detection (see
\S~2.2). Note that background spectra are taken from source-removed
event files to avoid contamination from nearby sources. The spectra
are fitted in the \textsc{xspec} (version 12.4.0,
\citealt{arnaud96,arnaud03}) software package using the C-statistic
\citep{cash79} due to the low photon counts in many of the spectra
(see Table~\ref{tab:xid}).   In order to improve the
counting statistics within each bin, we have re-binned the spectra to
fixed width spectral channels of $\sim$43.8 eV.

In fitting the X-ray spectra, we consider two different classes of
spectral models: (1) an intrinsically absorbed power-law, indicative
of AGN; and (2) a stellar model based on HMXB emission including
intrinsic absorption.  These models are designed to be simple, yet
physically meaningful, representations of the X-ray emission.  For
comparison with previous works, we also consider a simple power-law
with only Galactic absorption, represented by the \textsc{xspec} model
\textsc{pha(po)}, to measure the effective photon index
$\Gamma_{Eff}$.  As the C-statistic itself is not a measure of the
``goodness-of-fit'' (see, however, \citealt{lucy00}), we use the
\textsc{xspec} \textsc{goodness} command for comparing the different
spectral models (\S~3.1.3).  

\subsubsection{Model A: Absorbed Power-Law}

Our first model provides a simple parametrization of the X-ray
emission from an AGN, represented by a single power-law.  The model
includes the effects of both (Milky Way) Galactic and intrinsic
absorption and is represented by the \textsc{xspec} model
\textsc{pha(zpha(po))}. The X-ray spectra is thus defined by the
intrinsic absorption, N$_{\rm{H}}$, and photon index, $\Gamma$. As
these values can be strongly correlated for weak sources, we chose to
fix the photon index to $\Gamma=1.8$, typical for unobscured
AGNs \citep[i.e.][]{nandra96,tozzi06}.  The model (hereafter Model A)
thus represents a typical AGN and provides an estimate of
the level of obscuration present in our X-ray-identified SMGs.

\subsubsection{Model B: Absorbed HMXB}

\begin{figure}
\includegraphics[width=0.5\textwidth]{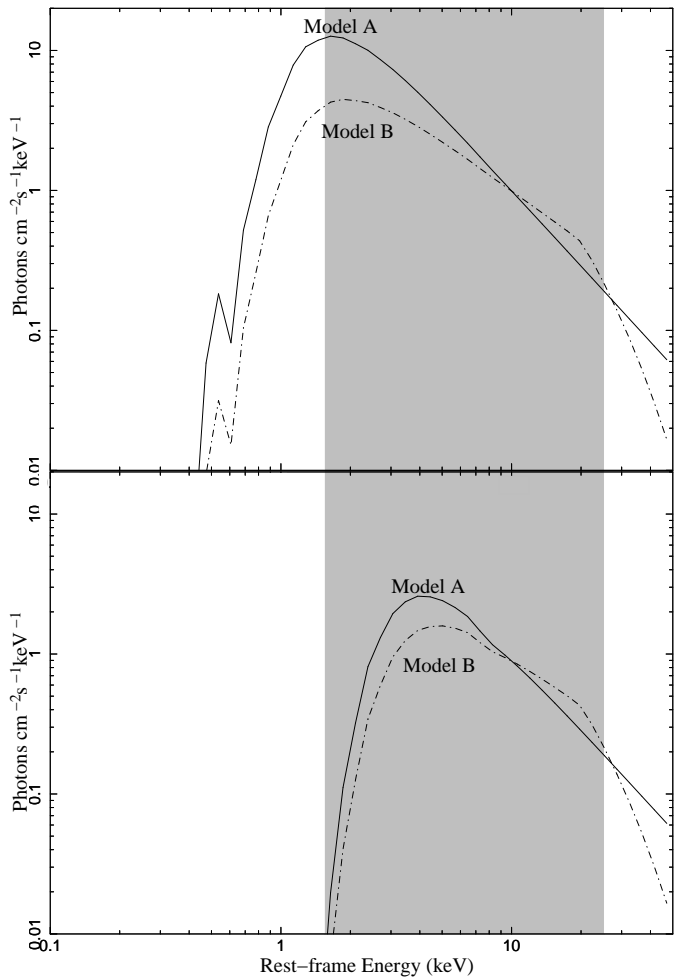}
\caption{Comparison of the X-ray spectral Models A (solid) and B
  (dot-dashed) normalized at $\sim$10 keV.  The models are shown for
  fiducial column densities of $10^{22}\rm~cm^{-2}$ (top)
  and $10^{23}\rm~cm^{-2}$ (bottom).  The shaded region indicates the
  effective rest-frame energies sampled by the 0.5-8.0 keV observed
  spectrum of a source at $z\sim2$.} 
\label{fig:xmodl}
\end{figure}

Our second model (Model B) is developed for emission due to star
formation and is based on the HMXB X-ray spectral model of
\cite{persic02}.  In summary, the X-ray emission from HMXBs can be
expressed as a broken power-law of the form 
\begin{equation}
f(\epsilon) = \begin{cases} \epsilon^{-\Gamma_{acc}}& \rm{if}
  \epsilon\le\epsilon_c\\  
\epsilon^{-\Gamma_{acc}}\rm{e}^{-[\epsilon-\epsilon_c]/\epsilon_F}& \rm{if}
  \epsilon>\epsilon_c  
\end{cases}
\end{equation}
\noindent where $\Gamma_{acc}$=1.2 \citep[typical of bright, accretion 
powered X-ray sources;][]{white93} with a cutoff energy of
$\epsilon_c\sim$20 keV and e-folding energy of $\epsilon_F\sim$12
keV. Ideally, when constructing a spectral model for stellar
processes, we should also include contributions from low mass X-ray
binaries (LMXBs) and supernovae.  However, supernovae contribute
little to $>2$keV rest-frame flux compared to HMXBs. While LMXBs may
contribute a considerable fraction of the hard X-ray flux, the low
mass stellar companion typically has not had time to evolve off the
main sequence and fill its Roche lobe by $z\sim$1-2.  For sources
in our sample with $z<1$, we may still use the HMXB-SFR relation as
\cite{persic07} showed that for moderate to high SFRs (SFRs$\gtrsim$50
M$_{\odot}$~yr$^{-1}$) the X-ray-SFR relation is similar to the
HMXB-SFR relation. Our stellar spectral model therefore consists of
only the HMXB emission, which is absorbed by both Galactic and intrinsic
material (\textsc{pha(zpha(hmxb))} in \textsc{xspec}, where the model
\textsc{hmxb} is defined as given above).  We include
intrinsic obscuration in Model B, since it is clear from
multi-wavelength evidence that SMGs are heavily dust-enshrouded
systems.  With $\Gamma_{acc}$ fixed to 1.2, we are left with only the
intrinsic obscuration and normalization to vary between spectra,
similar to Model A.

As shown in Figure~\ref{fig:xmodl}, there are immediate differences in
the spectral shapes of our adopted models.  Both models appear similar
at low energies; however, the difference in spectral slopes, as well
as the exponential cut-off in Model B, are apparent for higher
energies.  For high obscuration and low count spectra, it is
difficult to distinguish between Model A and B (\S~3.1.3). However,
the derived N$_{\rm{H}}$ values will vary according to the power-law
spectral slope.  Additionally, we can compare the X-ray-derived SFRs
of Model B with those obtained through our NIR-to-radio SED modeling
(\S~3.2). 

\subsubsection{Application of X-ray Spectral Models}

We now apply our set of spectral models to the X-ray identified AzTEC
SMGs.  To correctly fit the intrinsic absorption, which has a strong
energy dependence through the photo-electric cross-section, we require
accurate source redshift information.  This limits us to 32 out of our
original sample of 45 X-ray sources ($\sim$63 percent), including 5
sources in GOODS-N, 14 in GOODS-S, and 13 in COSMOS.  We favor the
spectroscopic redshift, whenever available, over the photometric
redshift.  Milky Way absorption values of 1.5, 0.9, and
2.5$\times$10$^{20}$ cm$^{-2}$ are included for the spectra, depending
on whether they were taken in GOODS-N, GOODS-S, or COSMOS,
respectively. The best-fit parameters for each set of models, as well
as their C-statistic values and associated rest-frame, absorption
corrected 2.0-10.0 keV luminosities, are given in
Table~\ref{tab:xspec}.  As a simple check, we have compared our
derived luminosities with those of previously published catalogs
\citep[i.e.][]{alex03,tozzi06} which correlate well with our results.   

In order to determine which of our sets of models offer the best fit
to the X-ray spectra, we run 2000 Monte Carlo simulations through the
\textsc{goodness} command in \textsc{xspec}, which provides the
percentage of simulations that have a C-statistic lower than the
observed spectrum.  The best-fit spectral models, the ones providing
the lowest goodness fraction, have been highlighted in boldface in
Table~\ref{tab:xspec}.  As one might expect, the models with the
lowest C-statistics tend to also provide the lowest goodness
fractions, indicating a very high probability that the observed
spectrum can be characterized by the best-fit model.  Models A and B
often show very similar C-statistics, which leads to only a few
percent difference in their goodness fractions.  These differences are
{\em not} statistically significant based on 10000 \textsc{fakeit}
simulated fits using an intrinsically absorbed $\Gamma$=1.8 power-law
as the template spectrum.

\begin{table*}
  \caption{X-ray spectral fits to identified AzTEC/X-ray sources.
  Spectral models used are: 
  Galactic dust- and intrinsically-absorbed AGN power-law
  (\textsc{pha(zpha(po))}, Model A) and Galactic dust- and
  intrinsically-absorbed power-law with an exponential cut-off
  relating to emission from HMXBs (\textsc{pha(zpha(hmxb))}, Model B).
  Models that offer the best fit to the X-ray spectra based on our
  simulations are emphasized in bold.  The relevant parameters given
  here are the intrinsic neutral hydrogen column density (N$_{\rm{H}}$
  in 10$^{22}\rm~cm^{-2}$); absorption corrected, rest-frame
  X-ray luminosity in the 2.0-10.0 keV energy band (L$_{\rm{X}}$ in
  10$^{43}\rm~ergs~s^{-1}$) and X-ray derived SFR (SFR$_{\rm{X}}$ in
  1000 M$_{\odot}$ yr$^{-1}$) for Model B assuming the P04
  relation. Errors are given at the 90 percent confidence level.}
  \label{tab:xspec}
  \begin{tabular}{@{}lcccccccc}
    \hline
\textit{Chandra} ID & $\Gamma_{Eff}$ & \multicolumn{3}{c}{Model A} & \multicolumn{4}{c}{Model B}\\
 & & N$_{\rm{H}}$ & L$_{\rm{X}}$ & C-stat & N$_{\rm{H}}$ &  L$_{\rm{X}}$ & SFR$_{\rm{X}}$ & C-stat\\
\hline
J123616.08+621514.1 & 0.98$_{-0.28}^{+ 0.23}$ &  16.46$_{ -6.20}^{+  7.64}$ & 4.40  & 159.3  &  \textbf{6.70$_{ -4.05}^{+  5.76}$} &\textbf{2.60}  & \textbf{26.0}  &\textbf{161.0}\\
J123635.86+620707.8 &-0.56$_{-0.45}^{+ 0.36}$ &  97.94$_{-29.10}^{+ 25.89}$ & 8.89  & 212.7  & \textbf{74.19$_{-27.01}^{+ 25.15}$} &\textbf{4.51}  & \textbf{45.1}  &\textbf{213.3}\\
J123711.32+621331.1 & 0.69$_{-0.55}^{+ 0.52}$ &   9.72$_{ -6.95}^{+ 21.27}$ & 0.92  & 189.0  &  $\mathbf{<16.28}$ &\textbf{0.62}  &  \textbf{6.2}  &\textbf{186.7}\\
J123711.98+621325.8 &-0.41$_{-0.61}^{+ 0.65}$ &  57.60$_{-24.30}^{+ 45.78}$ & 2.70  & 199.2  & \textbf{38.31$_{-18.49}^{+ 37.00}$} &\textbf{1.37}  & \textbf{13.7}  &\textbf{198.8}\\
J123716.63+621733.4 & 1.17$_{-0.06}^{+ 0.05}$ &   \textbf{2.10$_{ -0.23}^{+  0.25}$} & \textbf{9.95}  & \textbf{236.6}  &  0.52$_{ -0.17}^{+  0.18}$ &8.76  & 87.6  &239.4\\
J033158.25-274458.8 & 0.99$_{-0.55}^{+ 0.59}$ &   $<1.93$ & 0.07  & 177.3  &  $\mathbf{<0.83}$ &\textbf{0.08}  &  \textbf{0.8}  &\textbf{173.8}\\
J033204.48-274643.3 & 1.39$_{-0.20}^{+ 0.24}$ &   \textbf{1.00$_{ -0.82}^{+  1.02}$} & \textbf{1.08}  & \textbf{187.0}  &  $<0.37$ &1.07  & 10.7  &186.1\\
J033207.12-275128.6 & 0.74$_{-0.75}^{+-0.75}$ &   $\mathbf{<14.0}$ & \textbf{0.04}  & \textbf{189.1}  &  $<12.3$ &0.04  &  0.4  &189.2\\
J033209.71-274249.0 & 2.22$_{-0.28}^{+ 0.30}$ &   $\mathbf{<0.06}$ & \textbf{0.25}  & \textbf{199.4}  &  $<0.04$ &0.34  &  3.4  &238.4\\
J033212.23-274620.9 & 0.85$_{-0.13}^{+ 0.15}$ &   3.34$_{ -0.68}^{+  0.69}$ & 1.00  & 189.8  &  \textbf{1.53$_{ -0.50}^{+  0.58}$} &\textbf{0.88}  &  \textbf{8.8}  &\textbf{186.6}\\
J033215.32-275037.6 & 0.96$_{-0.29}^{+ 0.40}$ &   \textbf{0.82$_{ -0.42}^{+  0.41}$} & \textbf{0.01}  & \textbf{204.2}  &  0.34$_{ -0.31}^{+  0.35}$ &0.01  &  0.1  &205.7\\
J033222.17-274811.6 & 0.38$_{-0.28}^{+ 0.27}$ &  39.19$_{-10.40}^{+ 14.60}$ & 3.11  & 181.8  & \textbf{23.67$_{ -8.72}^{+ 10.86}$} &\textbf{1.66}  & \textbf{16.6}  &\textbf{182.1}\\
J033222.56-274815.0 &-0.43$_{-0.42}^{+ 0.49}$ &  94.11$_{-32.70}^{+ 55.73}$ & 3.51  & 182.9  & \textbf{55.83$_{-21.71}^{+ 45.48}$} &\textbf{1.49}  & \textbf{14.9}  &\textbf{180.6}\\
J033234.78-275534.0 & 1.06$_{-0.31}^{+ 0.28}$ &   0.43$_{ -0.19}^{+  0.27}$ & 4.0e-4& 167.9  &  $\mathbf{<0.45}$ &\textbf{5.0e-4}&\textbf{5.0e-3} &\textbf{167.7}\\
J033235.18-275215.7 & 0.64$_{-0.54}^{+ 0.39}$ &   \textbf{5.17$_{ -2.08}^{+  4.38}$} & \textbf{0.12}  & \textbf{179.7}  &  3.17$_{ -1.84}^{+  3.36}$ &0.10  &  1.0  &180.9\\
J033238.01-274401.2 & 1.77$_{-0.88}^{+ 1.00}$ &   $\mathbf{<5.53}$ & \textbf{0.14}  & \textbf{235.8}  &  $<3.95$ &0.14  &  1.4  &237.6\\
J033244.02-274635.9 & 2.01$_{-0.20}^{+ 0.20}$ &   $\mathbf{<0.96}$ & \textbf{3.64}  & \textbf{181.6}  &  $<0.26$ &3.15  & 31.5  &230.3\\
J033246.83-275120.9 & 0.95$_{-0.64}^{+ 0.52}$ &   4.32$_{ -3.16}^{+  4.92}$ & 0.20  & 209.2  &  $\mathbf{<5.64}$ &\textbf{0.17}  &  \textbf{1.7}  &\textbf{208.9}\\
J033302.94-275146.9 & 1.41$_{-0.26}^{+ 0.37}$ &  \textbf{10.36$_{ -6.34}^{+ 10.41}$} &\textbf{14.37}  & \textbf{175.5}  &  $<7.89$ &8.69  & 86.9  &182.4\\
J095905.05+022156.4 & 0.98$_{-1.25}^{+ 1.40}$ &  $\mathbf{<78.54}$ & \textbf{4.92}  &  \textbf{47.1}  & $<60.47$ &2.72  & 27.2  &47.6\\
J095929.70+021706.4 & 1.11$_{-1.23}^{+ 1.43}$ &   $<4.14$ & 0.83  &  92.7  &  $\mathbf{<3.28}$ &\textbf{0.93}  &  \textbf{9.3}  &\textbf{91.8}\\
J095945.15+023021.1 & 1.24$_{-1.38}^{+ 2.97}$ &   $<1.01$ & 0.41  & 140.0  &  $\mathbf{<0.95}$ &\textbf{0.57}  &  \textbf{5.7}  &\textbf{139.7}\\
J095953.85+021853.6 & 0.57$_{-0.59}^{+ 0.58}$ &   5.56$_{ -3.14}^{+  3.98}$ & 0.79  & 103.2  &  \textbf{3.27$_{ -2.56}^{+  3.66}$} &\textbf{0.66}  &  \textbf{6.6}  &\textbf{103.7}\\
J095959.96+020633.1 & 0.52$_{-0.74}^{+ 0.66}$ &   \textbf{5.53$_{ -3.11}^{+  4.30}$} & \textbf{0.39}  &  \textbf{79.0}  &  3.71$_{ -2.65}^{+  3.92}$ &0.33  &  3.3  &80.0\\
J100003.73+020206.4 & 1.00$_{-1.53}^{+ 2.06}$ &  $\mathbf{<124.89}$ & \textbf{6.16}  & \textbf{141.2}  &  $<82.08$ &2.52  & 25.2  &140.8\\
J100006.11+015239.2 & 1.77$_{-0.57}^{+ 0.70}$ &  $\mathbf{<1.48}$ & \textbf{2.25}  & \textbf{115.6}  &  $<0.83$ &2.26  & 22.6  &118.3\\
J100006.55+023259.3 & 1.26$_{-0.54}^{+ 0.58}$ &  $\mathbf{<2.98}$ & \textbf{0.82}  &  \textbf{87.1}  &  $<1.56$ &0.79  &  7.9  &86.4\\
J100033.61+014902.0 & 1.57$_{-0.45}^{+ 0.50}$ &   $<0.68$ & 0.75  & 104.8  &  $\mathbf{<0.29}$ &\textbf{0.95}  &  \textbf{9.5}  &\textbf{107.5}\\
J100055.34+023441.1 & 1.85$_{-0.19}^{+ 0.19}$ &   $\mathbf{<0.44}$ &\textbf{24.88}  & \textbf{159.2}  &  $<0.10$ &26.84 &268.4  &181.6\\
J100107.46+015718.1 & 1.59$_{-0.60}^{+ 0.79}$ &   0.88$_{ -0.86}^{+  2.27}$ & 1.46  & 148.5  &  $\mathbf{<1.50}$ &\textbf{1.49}  & \textbf{14.9}  &\textbf{149.9}\\
J100116.15+023606.9 & 1.72$_{-0.56}^{+ 0.60}$ &   \textbf{0.20$_{ -0.19}^{+  1.31}$} & \textbf{5.84}  & \textbf{100.3}  & $<0.69$ &6.72  &  67.1  &102.6\\
J100139.73+022548.5 & 3.23$_{-0.71}^{+ 0.79}$ &   $\mathbf{<0.05}$ & \textbf{0.01}  &  \textbf{81.3}  &  $<0.04$ &0.02  &  0.2  &97.5\\
\hline
\end{tabular}
\end{table*}

We find that $\sim$53 percent (17/32) of the AzTEC/X-ray sources have
X-ray spectra that immediately favors an AGN origin.  Of these,
$\sim$70 percent show evidence for heavy obscuration with N$_{\rm{H}}\gtrsim
10^{23}\rm~cm^{-2}$.  Regardless of the best-fit spectral model, the
majority of AzTEC/X-ray sources (22/32) have 2.0 to 10.0 keV
rest-frame luminosities of $\gtrsim$10$^{43}\rm~ergs~s^{-1}$, heavily
favoring an AGN interpretation.  Note that the derived luminosities
are sensitive to the choice of the X-ray model.  For those AzTEC/X-ray
sources that favor the starburst model Model B, we use the X-ray
luminosity to SFR relation of P04 to estimate a SFR, assuming no
contribution from a buried AGN.  There is some uncertainty in the
exact form of the X-ray-to-SFR scaling relation as discussed by
\cite{mineo11}; however, many of these relations consider local, low
SFR ($\lesssim 10\rm~M_{\odot}~yr^{-1}$) sources during their
construction.  As we are concerned with potentially high SFRs, we
favor the P04 and \cite{persic07} SFR-X-ray scaling relations; using the
\cite{ranalli03} relation, or similar, would decrease the estimated
SFRs by a factor of $\sim$2-5. The high X-ray luminosities would
require very strong SFRs on the order of $\gtrsim
10^3-10^4\rm~M_{\odot}~yr^{-1}$, which is pushing the limits for
typical SMGs.  However, there are 5 sources with L$_{\rm{X}}\lesssim
10^{42}\rm~ergs~s^{-1}$ which are candidates to be starburst powered 
X-ray sources.  These sources account for $\sim$16 percent of our
X-ray-identified SMG sample; consistent with the
starburst-powered fraction of LNPS10 ($\sim$17$\pm$6 percent). We
caution, however, that this does not necessarily imply that their
X-ray emission is dominated by star formation (see \S~3.2.2,
Tables~\ref{tab:sedas}).  These results thus show that the
bulk of the X-ray emission from our SMG sample is predominately
produced by obscured AGNs.  

\subsection{NIR-to-Radio SED Modeling}

For an alternative view of the AGN and star formation contributions,
we now examine the near-IR-to-radio SEDs of the AzTEC/X-ray sources.
To be luminous at (sub-)millimetre wavelengths, a source must contain
dust heated to T$\sim$30K \citep{chapman05,pope06} through some
central engine.  While it is possible to have (sub-)mm emission due to
synchrotron processes from radio-loud AGN \citep[e.g.,][]{vieira10},
the corresponding radio fluxes would have to be significantly larger
\citep[on order 1-100 mJy;][]{dezotti10,vieira10} than those observed
for our AzTEC/X-ray sources, which range from 0.02 to 0.65 mJy
(Table~\ref{tab:mid}).  The required dust heating must then be
accomplished either by star formation, AGN activity, or some
combination of the two.  

For our SED modeling, we consider the templates of \cite{efstathiou00}
and \cite{siebenmorgen04} to parametrize emission from a
starburst (SB) and AGN component, respectively.  This selection of
templates is widely used in the literature and has shown to provide
reasonable results to similar classes of sources over the NIR-to-mm
wavelength regime \citep[i.e.][and references
  therein]{efstathiou00,siebenmorgen04,meng10,serra11,younger11,yun12}.
For this work, we favor the \cite{siebenmorgen04} AGN models as opposed
to torus models as we are more interested in the integrated AGN host
properties rather than the centralized nuclear region.  Additionally,
these models are built from basic radiative transfer models,
incorporating relevant dust emission/absorption physics, with simple
parametrizations comparable to the SB models.

In order to estimate the total SED, we apply a simple linear
combination of the two template sets.  Since this approach may
introduce strong template-parameter degeneracies into our summed
SEDs, we use a Monte Carlo Markov Chain (MCMC) technique to perform the
fitting.  While computationally slower compared to direct maximum
likelihood (least squares) fitting, MCMC has the advantage that the
full set of posterior parameter distributions are returned - allowing
for direct inspection of the posteriors for degeneracies that may bias
our interpretations of the fits (see Figure~\ref{fig:sedcont}).  The full
details of this method will be presented in Johnson et al. (in prep).
Here, we briefly describe the adopted models and their implications on
the AzTEC/X-ray source population. 

\subsubsection{SED Models and Fitting}

Before applying the SED templates to the observed SEDs, it is helpful
to have an understanding of how the templates parametrize the
underlying physics and resulting IR emission.  In the
\cite{efstathiou00} templates, emission from a dusty starburst is
traced from a single star forming GMC with the cloud
optical depth ($\tau_{\nu}$) and starburst age setting the overall
shape of the SED.  Specifically, $\tau_{\nu}$ controls the strength of
the PAH and silicate features, while older starburst ages shift the IR
peak to longer wavelengths.  A normalization factor is then required
to scale the emission from a single GMC to the full system.  This
normalization is comparable to the SFR at the onset of the burst as
\cite{efstathiou00} assume an exponentially decaying SFR history of
the form
\begin{equation}
SFR(t)\approx SFR(0)\rm{e}^{-t/20\rm~Myr}
\end{equation}
where $t$ is the SB age; SFR estimates obtained this way are
approximately 2-3 times lower than more traditional FIR SFR indicators
-- for example, the \cite{kennicutt98} relation. The AGN models
are described by a single central illuminating source with intrinsic
luminosity L surrounded by a spherical dust distribution of size R and
the visual extinction (A$_V$).  The dust distribution, temperature,
strength of absorption/emission lines, etc. are adjusted through a
combination of the size and visual extinction.  It should be noted
that the \cite{siebenmorgen04} AGN templates make a number of
simplifications compared to alternative AGN models.  Modern AGN
templates \citep[e.g.][and references therein]{fritz06,nenkova08}
consider the AGN to be surrounded by torus, generally composed of a
clumpy material, whose geometry flares outward.  This geometry
naturally falls in line with the standard AGN unified model where
looking through the torus results in Type 2 (obscured) AGN while Type
1 (unobscured) AGN are produced from 'face-on' observations. The
\cite{siebenmorgen04} models obviously lack the asymmetry and clumpy
distribution of the traditional AGN torus but are able to recreate the
same effects; Siebenmorgen et al. comments that it is the dust mass
and distance from the source (set by A$_V$ and R) that are most
important.  Though torus geometries may extend to the kpc scale
\cite[e.g.][]{granato94,fritz06,nenkova08} and can produce significant
cold dust emission, they lack the dust intrinsic to the host galaxy,
whose geometry extends well beyond that of a nuclear torus.  Given
that the photometry for our sample can not resolve our sources, we
believe the \cite{siebenmorgen04} models to be better representative
of galactic emission resulting from an AGN than the traditional torus
models. 

Using the above models, we have three to six free parameters with
6-7 available SED data points per AzTEC/X-ray source.  In our fitting,
we are able to predict an X-ray luminosity from the FIR luminosity/SFR
using the relations of \cite{marconi04} and P04 for the AGN and SB
models, respectively. This allows us to then use the observed X-ray
luminosities in Table~\ref{tab:xspec} as an additional prior to the
fits.  As the SB models do not account for any radio emission, we
employ the radio-FIR correlation of \cite{yun01} to add a radio 'tail'
to the templates. Note, however, that this may still pose some
uncertainty when combining templates as there is scatter in this
relationship \citep[e.g.][]{carilli00,chapman05} and it does not
predict any radio emission resulting from an AGN. 

In fitting the near-IR-to-radio SEDs of our X-ray-detected SMGs, we
consider two combinations of the SED templates: (1) AGN and SB
templates including the observed X-ray luminosity and X-ray-absorbing
column density as priors to the AGN luminosity, SB SFR and AGN
A$_{\rm{V}}$ and (2) SB only without the additional X-ray
constraints. The first set of models serves to estimate the AGN
contribution to the bolometric and 1.1 mm emission.  The SB only fits
provide a measure of the necessity of the AGN templates.  The X-ray
luminosity prior had to be excluded for these fits as their inclusion
produced unreasonable results (see \S~3.2.2).  Tables~\ref{tab:sedas}
and \ref{tab:seds} and Figure~\ref{fig:seds} show the results of our
MCMC fitting technique to our AzTEC/X-ray sample.  For each set of
best-fit parameters, we calculate the log of the likelihood, ln(L);
higher values of ln(L) indicate a higher probability that the data is
consistent with the best-fit model.   

\begin{table*}
  \caption{Best-fit parameters for the composite AGN+SB models based
    on the broadband photometry of AzTEC/X-ray sources.  The
    predicted X-ray luminosities are compared to those derived from
    the X-ray spectral modeling (\S~3.1.4) to provide additional
    weights in calculating the likelihoods.  Errors are given at the
    1$\sigma$ confidence level after marginalizing over all other free
    parameters in the fitted templates.  Col.(1): \textit{Chandra} Source ID.
    Col.(2), (3), and (4): AGN template galaxy outer radius, intrinsic
    luminosity and visual extinction.  Col.(5), (6), and (7): SB
    template normalization, age, and optical depth.}
  \label{tab:sedas}
  \begin{tabular}{@{}ccccccc}
    \hline
    & \multicolumn{3}{c}{AGN} & \multicolumn{3}{c}{SB}\\
    \textit{Chandra} ID & R & L & A$_{V}$ & Norm & Age & $\tau_{\nu}$\\
    & kpc & log(L$_{\odot}$) & mag & & Myr &\\
    (1) &(2)&(3)&(4)&(5)&(6)&(7)\\
    \hline
    J123616.08+621514.1 & 0.13$^{+0.15}_{-0.01}$ & 11.51$^{+0.05}_{-0.18}$ & 26.10$^{+36.73}_{-17.57}$ & 8532.40$^{+650.39}_{-490.44}$ & 44.95$^{+11.53}_{-7.68}$ & 192.73$^{+7.26}_{-41.29}$ \\
    J123635.86+620707.8 & 14.30$^{+1.70}_{-14.17}$ & 10.52$^{+0.11}_{-0.18}$ & 125.85$^{+2.14}_{-124.84}$ & 213.66$^{+114.63}_{-92.54}$ & 56.24$^{+15.75}_{-28.36}$ & 196.40$^{+3.60}_{-139.53}$ \\
    J123711.32+621331.1 & 0.91$^{+15.07}_{-0.79}$ & 10.41$^{+0.11}_{-0.12}$ & 56.00$^{+31.97}_{-54.99}$ & 7965.80$^{+182.33}_{-211.12}$ & 56.66$^{+6.41}_{-10.49}$ & 198.62$^{+1.37}_{-43.97}$ \\
    J123711.98+621325.8 & 9.20$^{+6.74}_{-9.08}$ & 10.81$^{+0.13}_{-0.14}$ & 74.24$^{+53.54}_{-73.24}$ & 2685.09$^{+113.89}_{-287.82}$ & 43.46$^{+12.53}_{-5.76}$ & 195.49$^{+4.51}_{-42.44}$ \\
    J123716.63+621733.4 & 15.98$^{+0.02}_{-7.55}$ & 12.25$^{+0.02}_{-0.01}$ & 2.10$^{+1.80}_{-1.02}$ & 9451.94$^{+241.06}_{-179.17}$ & 1.37$^{+4.85}_{-1.37}$ & 151.18$^{+46.25}_{-47.75}$ \\
    J033158.25-274458.8 & 14.26$^{+1.74}_{-14.13}$ & 9.21$^{+0.20}_{-0.13}$ & 1.00$^{+3.91}_{-0.01}$ & 1097.69$^{+2.81}_{-4.57}$ & 70.67$^{+1.33}_{-6.40}$ & 199.88$^{+0.11}_{-48.26}$ \\
    J033204.48-274643.3 & 14.63$^{+1.36}_{-14.49}$ & 10.67$^{+0.16}_{-0.19}$ & 1.03$^{+9.78}_{-0.03}$ & 233.46$^{+96.74}_{-53.53}$ & 68.27$^{+3.73}_{-10.70}$ & 105.65$^{+94.31}_{-55.34}$ \\
    J033207.12-275128.6 & 8.48$^{+7.51}_{-8.35}$ & 8.97$^{+0.11}_{-0.13}$ & 67.61$^{+7.31}_{-66.61}$ & 246.75$^{+13.02}_{-24.16}$ & 71.54$^{+0.45}_{-13.85}$ & 197.10$^{+2.88}_{-44.25}$ \\
    J033209.71-274249.0 & 15.91$^{+0.09}_{-5.84}$ & 11.00$^{+0.01}_{-0.01}$ & 1.00$^{+0.01}_{-0.01}$ & 2040.88$^{+3.22}_{-3.52}$ & 71.22$^{+0.78}_{-0.50}$ & 198.51$^{+1.48}_{-31.29}$ \\
    J033212.23-274620.9 & 15.91$^{+0.09}_{-7.64}$ & 11.25$^{+0.01}_{-0.01}$ & 1.08$^{+0.89}_{-0.08}$ & 2106.95$^{+8.74}_{-6.94}$ & 71.46$^{+0.54}_{-7.25}$ & 193.54$^{+6.45}_{-42.36}$ \\
    J033215.32-275037.6 & 5.65$^{+10.34}_{-5.53}$ & 10.70$^{+0.05}_{-0.29}$ & 1.03$^{+1.99}_{-0.03}$ & 3280.00$^{+23.15}_{-27.60}$ & 70.60$^{+1.40}_{-6.14}$ & 50.31$^{+47.39}_{-0.30}$ \\
    J033222.17-274811.6 & 15.38$^{+0.62}_{-15.11}$ & 11.27$^{+0.07}_{-0.02}$ & 4.46$^{+10.92}_{-2.38}$ & 2483.75$^{+80.91}_{-22.82}$ & 44.38$^{+11.97}_{-6.94}$ & 53.72$^{+43.88}_{-3.72}$ \\
    J033222.56-274815.0 & 13.81$^{+2.19}_{-13.68}$ & 10.85$^{+0.19}_{-0.08}$ & 1.08$^{+2.24}_{-0.08}$ & 3844.91$^{+37.39}_{-72.25}$ & 71.97$^{+0.03}_{-7.59}$ & 57.91$^{+40.71}_{-7.88}$ \\
    J033235.18-275215.7 & 0.15$^{+0.10}_{-0.02}$ & 9.76$^{+0.02}_{-0.01}$ & 50.82$^{+1.18}_{-18.32}$ & 109.77$^{+5.04}_{-6.44}$ & 71.98$^{+0.02}_{-7.41}$ & 198.63$^{+1.37}_{-46.96}$ \\
    J033238.01-274401.2 & 15.69$^{+0.31}_{-15.56}$ & 9.51$^{+0.07}_{-0.15}$ & 10.90$^{+19.09}_{-9.90}$ & 1019.35$^{+22.35}_{-21.65}$ & 36.06$^{+8.06}_{-8.69}$ & 198.61$^{+1.39}_{-43.73}$ \\
    J033244.02-274635.9 & 0.13$^{+0.11}_{-0.01}$ & 11.51$^{+0.04}_{-0.01}$ & 5.92$^{+0.08}_{-1.70}$ & 2523.09$^{+38.84}_{-38.49}$ & 57.21$^{+6.09}_{-11.43}$ & 101.31$^{+45.07}_{-46.64}$ \\
    J033246.83-275120.9 & 0.14$^{+0.11}_{-0.01}$ & 10.00$^{+0.02}_{-0.01}$ & 30.97$^{+0.03}_{-14.38}$ & 868.22$^{+9.67}_{-7.11}$ & 71.34$^{+0.66}_{-6.77}$ & 197.29$^{+2.71}_{-45.44}$ \\
    J033302.94-275146.9 & 0.25$^{+0.22}_{-0.01}$ & 12.26$^{+0.03}_{-0.01}$ & 6.96$^{+8.01}_{-2.80}$ & 7001.07$^{+139.78}_{-215.29}$ & 44.91$^{+11.39}_{-7.35}$ & 100.74$^{+45.91}_{-46.93}$ \\
    J095905.05+022156.4 & 2.39$^{+13.57}_{-2.26}$ & 11.40$^{+0.13}_{-0.13}$ & 7.42$^{+120.08}_{-6.42}$ & 8693.26$^{+308.01}_{-550.64}$ & 44.42$^{+11.06}_{-6.48}$ & 151.03$^{+44.55}_{-47.51}$ \\
    J095929.70+021706.4 & 6.22$^{+9.78}_{-6.09}$ & 10.61$^{+0.12}_{-0.13}$ & 16.33$^{+1.22}_{-15.32}$ & 4819.83$^{+39.78}_{-238.62}$ & 71.79$^{+0.21}_{-12.12}$ & 148.38$^{+48.37}_{-44.31}$ \\
    J095945.15+023021.1 & 9.82$^{+6.00}_{-8.78}$ & 10.75$^{+0.01}_{-0.01}$ & 1.31$^{+0.68}_{-0.31}$ & 1797.01$^{+8.54}_{-19.38}$ & 71.87$^{+0.13}_{-7.66}$ & 60.11$^{+38.85}_{-10.11}$ \\
    J095953.85+021853.6 & 0.13$^{+0.11}_{-0.01}$ & 10.52$^{+0.08}_{-0.01}$ & 31.08$^{+5.99}_{-14.31}$ & 1677.68$^{+15.21}_{-14.04}$ & 71.88$^{+0.12}_{-7.63}$ & 195.74$^{+4.25}_{-44.40}$ \\
    J095959.96+020633.1 & 12.49$^{+3.50}_{-12.37}$ & 10.09$^{+0.07}_{-0.02}$ & 52.38$^{+0.20}_{-33.85}$ & 669.15$^{+7.07}_{-11.74}$ & 46.63$^{+8.96}_{-9.36}$ & 190.37$^{+9.63}_{-39.41}$ \\
    J100003.73+020206.4 & 12.25$^{+3.75}_{-12.12}$ & 11.18$^{+0.11}_{-0.17}$ & 121.10$^{+6.90}_{-120.08}$ & 995.74$^{+44.76}_{-169.08}$ & 70.55$^{+1.45}_{-24.73}$ & 54.92$^{+43.17}_{-4.92}$ \\
    J100006.11+015239.2 & 12.59$^{+3.41}_{-12.47}$ & 11.07$^{+0.13}_{-0.16}$ & 7.90$^{+0.02}_{-6.89}$ & 2859.00$^{+34.58}_{-188.34}$ & 71.82$^{+0.18}_{-13.53}$ & 185.32$^{+14.67}_{-33.72}$ \\
    J100006.55+023259.3 & 0.92$^{+2.81}_{-0.75}$ & 10.43$^{+0.07}_{-0.11}$ & 1.01$^{+0.93}_{-0.01}$ & 1905.42$^{+19.50}_{-16.07}$ & 71.04$^{+0.96}_{-6.52}$ & 194.34$^{+5.66}_{-42.19}$ \\
    J100033.61+014902.0 & 6.84$^{+9.16}_{-5.87}$ & 10.64$^{+0.13}_{-0.12}$ & 1.47$^{+0.09}_{-0.34}$ & 2418.84$^{+26.17}_{-41.92}$ & 71.81$^{+0.19}_{-7.48}$ & 193.21$^{+6.79}_{-41.24}$ \\
    J100055.34+023441.1 & 15.73$^{+0.26}_{-7.54}$ & 12.75$^{+0.02}_{-0.01}$ & 2.32$^{+0.03}_{-0.31}$ & 1366.06$^{+69.89}_{-49.88}$ & 17.24$^{+8.42}_{-6.91}$ & 54.57$^{+43.64}_{-4.57}$ \\
    J100107.46+015718.1 & 15.60$^{+0.40}_{-12.86}$ & 10.87$^{+0.25}_{-0.08}$ & 2.43$^{+5.60}_{-1.42}$ & 2202.64$^{+101.75}_{-130.42}$ & 70.94$^{+1.06}_{-6.85}$ & 197.34$^{+2.66}_{-46.46}$ \\
    J100116.15+023606.9 & 0.13$^{+0.11}_{-0.01}$ & 11.82$^{+0.11}_{-0.15}$ & 8.08$^{+0.01}_{-0.07}$ & 4426.14$^{+18.13}_{-15.78}$ & 64.22$^{+7.29}_{-6.83}$ & 196.37$^{+3.62}_{-43.90}$ \\
    J100139.73+022548.5 & 0.99$^{+0.32}_{-0.48}$ & 9.00$^{+0.01}_{-0.01}$ & 1.00$^{+0.01}_{-0.01}$ & 10.41$^{+0.08}_{-0.09}$ & 2.07$^{+4.26}_{-2.07}$ & 51.98$^{+44.46}_{-1.96}$\\
     \hline
  \end{tabular}
\end{table*}

\begin{table*}
  \caption{Continuation of Table~\ref{tab:sedas} containing the
    derived physical properties based on the fitted parameters.
    Col.(1): \textit{Chandra} Source ID. Col.(2):  Total
    model-derived, rest-frame bolometric IR luminosity from the AGN
    and SB templates over the wavelength range
    $\sim$0.001-1500$\mu$m. Col.(3): SFR derived from Eqn.~2
    (\S~3.2.1) and Cols. 5 \& 6 of Table~\ref{tab:sedas}.  Col.(4):
    SED derived X-ray luminosity. Col.(5): Fractional contribution of
    AGN template to total model emission at 1.1mm. Col.(6):
    Fractional contribution of model to observed 1.1mm flux. Col.(7):
    ln(L) of best-fit parameters.} 
  \begin{tabular}{@{}ccccccc}
    \hline
    & IR & & X-ray & & \\
    \textit{Chandra} ID &  Lum. & SFR & Lum. & f$_{AGN,1.1mm}$ & f$_{1.1mm}$ & ln(L)\\
    & 10$^{12}$L$_{\odot}$ & M$_{\odot}$ yr$^{-1}$ & 10$^{43}$ergs s$^{-1}$ &  & &\\
    (1) &(2)&(3)&(4)&(5)&(6)&(7)\\
    \hline
    J123616.08+621514.1 & 11.10$^{+3.30}_{-3.20}$ & 901 & 4.84$^{+0.57}_{-1.26}$ & 0.00 & 0.001 & -27.94\\
    J123635.86+620707.8 & 0.23$^{+0.33}_{-0.11}$ & 12 & 0.79$^{+0.19}_{-0.21}$ & 0.91 & 0.312 & -11.40\\
    J123711.32+621331.1 & 7.10$^{+2.61}_{-1.21}$ & 468 & 0.72$^{+0.13}_{-0.14}$ & 0.01 & 0.005 & -40.04\\
    J123711.98+621325.8 & 3.64$^{+0.65}_{-1.22}$ & 305 & 1.39$^{+0.34}_{-0.31}$ & 0.16 & 0.059 & -9.59\\
    J123716.63+621733.4 & 13.00$^{+15.90}_{-0.20}$ & 8825 & 17.30$^{+0.63}_{-0.11}$ & 0.23 & 0.131 & -2868.31\\
    J033158.25-274458.8 & 0.67$^{+0.12}_{-0.03}$ & 32 & 0.07$^{+0.03}_{-0.01}$ & 0.05 & 0.017 & -2226.84\\
    J033204.48-274643.3 & 0.20$^{+0.07}_{-0.06}$ & 7 & 1.05$^{+0.35}_{-0.33}$ & 0.44 & 0.021 & -24.87\\
    J033207.12-275128.6 & 0.15$^{+0.06}_{-0.01}$ & 6 & 0.04$^{+0.01}_{-0.01}$ & 0.25 & 0.010 & -23.65\\
    J033209.71-274249.0 & 1.32$^{+0.02}_{-0.03}$ & 57 & 1.87$^{+0.01}_{-0.01}$ & 0.10 & 0.070 & -135568.62\\
    J033212.23-274620.9 & 1.43$^{+0.26}_{-0.02}$ & 59 & 2.94$^{+0.02}_{-0.01}$ & 0.11 & 0.038 & -3350.01\\
    J033215.32-275037.6 & 2.04$^{+0.34}_{-0.10}$ & 96 & 1.10$^{+0.09}_{-0.44}$ & 0.01 & 0.004 & -1280.28\\
    J033222.17-274811.6 & 3.39$^{+0.78}_{-0.97}$ & 269 & 3.08$^{+0.46}_{-0.13}$ & 0.18 & 0.098 & -60.08\\
    J033222.56-274815.0 & 2.32$^{+0.51}_{-0.01}$ & 105 & 1.48$^{+0.56}_{-0.22}$ & 0.04 & 0.022 & -62.72\\
    J033235.18-275215.7 & 0.07$^{+0.02}_{-0.01}$ & 3 & 0.18$^{+0.01}_{-0.01}$ & 0.02 & 0.000 & -302.13\\
    J033238.01-274401.2 & 1.70$^{+0.50}_{-0.38}$ & 167 & 0.13$^{+0.02}_{-0.03}$ & 0.10 & 0.024 & -111.65\\
    J033244.02-274635.9 & 2.53$^{+0.92}_{-0.35}$ & 144 & 4.75$^{+0.44}_{-0.16}$ & 0.00 & 0.000 & -89.91\\
    J033246.83-275120.9 & 0.53$^{+0.10}_{-0.01}$ & 24 & 0.30$^{+0.01}_{-0.01}$ & 0.00 & 0.000 & -4001.74\\
    J033302.94-275146.9 & 10.70$^{+2.30}_{-2.60}$ & 741 & 17.39$^{+1.32}_{-0.25}$ & 0.00 & 0.001 & -61.65\\
    J095905.05+022156.4 & 11.40$^{+2.60}_{-3.30}$ & 942 & 4.08$^{+0.93}_{-0.88}$ & 0.01 & 0.009 & -4.53\\
    J095929.70+021706.4 & 2.87$^{+0.94}_{-0.06}$ & 133 & 0.97$^{+0.21}_{-0.22}$ & 0.06 & 0.027 & -2.33\\
    J095945.15+023021.1 & 1.11$^{+0.23}_{-0.01}$ & 49 & 1.19$^{+0.04}_{-0.01}$ & 0.04 & 0.007 & -588.57\\
    J095953.85+021853.6 & 1.02$^{+0.22}_{-0.01}$ & 46 & 0.78$^{+0.15}_{-0.03}$ & 0.00 & 0.000 & -222.44\\
    J095959.96+020633.1 & 0.82$^{+0.27}_{-0.19}$ & 65 & 0.36$^{+0.05}_{-0.02}$ & 0.51 & 0.085 & -174.35\\
    J100003.73+020206.4 & 0.76$^{+0.56}_{-0.11}$ & 29 & 2.65$^{+0.58}_{-0.74}$ & 0.75 & 0.212 & -18.55\\
    J100006.11+015239.2 & 1.80$^{+0.64}_{-0.09}$ & 78 & 2.21$^{+0.57}_{-0.56}$ & 0.21 & 0.063 & -11.98\\
    J100006.55+023259.3 & 1.17$^{+0.21}_{-0.04}$ & 54 & 0.68$^{+0.07}_{-0.13}$ & 0.00 & 0.000 & -4319.74\\
    J100033.61+014902.0 & 1.47$^{+0.31}_{-0.03}$ & 66 & 1.00$^{+0.23}_{-0.21}$ & 0.03 & 0.005 & -323.60\\
    J100055.34+023441.1 & 9.53$^{+0.69}_{-0.73}$ & 576 & 40.84$^{+1.98}_{-0.37}$ & 0.47 & 0.076 & -18184.24\\
    J100107.46+015718.1 & 1.40$^{+0.31}_{-0.06}$ & 63 & 1.52$^{+0.86}_{-0.23}$ & 0.12 & 0.028 & -93.88\\
    J100116.15+023606.9 & 3.85$^{+0.71}_{-0.63}$ & 178 & 8.23$^{+1.68}_{-1.96}$ & 0.00 & 0.000 & -1096.50\\
    J100139.73+022548.5 & 0.01$^{+0.02}_{-0.01}$ & 9 & 0.04$^{+0.01}_{-0.01}$ & 0.32 & 0.001 & -187114.26\\
     \hline
  \end{tabular}
\end{table*}

\begin{table*}
  \caption{Best-fit SED parameters using only the SB models.  The
    SED derived X-ray luminosity is left as a free parameter and
    provides no additional constraint to the SED fitting.  Errors are
    given at the 1$\sigma$ confidence level after marginalizing over
    all other free parameters in the fitted templates. Col.(1):
    \textit{Chandra} Source ID. Col.(2), (3), and (4): SB template
    normalization, age, and optical depth. Col.(5): Total
    model-derived, rest-frame bolometric IR luminosity from
    $\sim$0.001-1500$\mu$m. Col.(6): SFR derived from Eqn.~2
    (\S~3.2.1) and Cols. 2 \& 3. Col.(7): SED derived X-ray
    luminosity. Col.(8): Fractional contribution of model to observed
    1.1mm flux. Col.(9): ln(L) of best-fit parameters.}
  \label{tab:seds}
  \begin{tabular}{@{}ccccccccc}
    \hline
    \textit{Chandra} ID & Norm & AGE & $\tau_{\nu}$ & IR Lum. & SFR & X-ray Lum. & f$_{1.1mm}$ & ln(L)\\
    & & Myr & & 10$^{12}$L$_{\odot}$ & M$_{\odot}$ yr$^{-1}$ & 10$^{43}\rm~ergs~s^{-1}$ &\\
    (1)&(2)&(3)&(4)&(5)&(6)&(7)&(8)&(9)\\
    \hline
    J123616.08+621514.1 & 7874.8$^{+261.8}_{-247.1}$ & 45.2$^{+9.8}_{-6.9}$ & 150.0$^{+41.5}_{-41.5}$ & 9.87$^{+2.43}_{-2.47}$ & 822 & 0.10$^{+0.03}_{-0.03}$ & 1.22 & -48.42\\
    J123635.86+620707.8 & 246.9$^{+88.3}_{-81.7}$ & 45.2$^{+26.8}_{-12.9}$ & 199.4$^{+0.6}_{-88.3}$ & 0.31$^{+0.19}_{-0.19}$ & 26 & 0.00$^{+0.01}_{-0.01}$ & 0.04 & -12.25\\
    J123711.32+621331.1 & 8412.4$^{+156.2}_{-184.8}$ & 56.9$^{+6.1}_{-9.9}$ & 199.0$^{+1.0}_{-41.1}$ & 7.41$^{+2.63}_{-1.17}$ & 489 & 0.08$^{+0.03}_{-0.01}$ & 0.81 & -28.15\\
    J123711.98+621325.8 & 2832.8$^{+138.5}_{-625.4}$ & 45.2$^{+10.0}_{-7.0}$ & 199.6$^{+0.4}_{-48.2}$ & 3.55$^{+0.91}_{-0.92}$ & 296 & 0.04$^{+0.01}_{-0.01}$ & 0.30 & -9.35\\
    J123716.63+621733.4 & 6318.3$^{+67.8}_{-66.2}$ & 36.8$^{+6.9}_{-9.2}$ & 200.0$^{+0.0}_{-42.2}$ & 10.30$^{+3.30}_{-1.90}$ & 1003 & 0.12$^{+0.04}_{-0.03}$ & 1.14 & -6057.59\\
    J033158.25-274458.8 & 1123.2$^{+3.4}_{-2.3}$ & 72.0$^{+0.0}_{-6.9}$ & 199.8$^{+0.2}_{-43.2}$ & 0.66$^{+0.13}_{-0.01}$ & 31 & 0.01$^{+0.01}_{-0.01}$ & 0.29 & -2217.17\\
    J033204.48-274643.3 & 302.8$^{+57.6}_{-62.5}$ & 71.9$^{+0.1}_{-9.7}$ & 101.1$^{+86.4}_{-51.1}$ & 0.18$^{+0.06}_{-0.04}$ & 8 & 0.00$^{+0.01}_{-0.01}$ & 0.03 & -26.34\\
    J033207.12-275128.6 & 253.3$^{+14.9}_{-20.3}$ & 71.8$^{+0.2}_{-12.9}$ & 198.9$^{+1.1}_{-45.1}$ & 0.15$^{+0.06}_{-0.01}$ & 7 & 0.00$^{+0.01}_{-0.01}$ & 0.03 & -23.76\\
    J033209.71-274249.0 & 2277.2$^{+3.4}_{-3.4}$ & 71.9$^{+0.1}_{-6.8}$ & 200.0$^{+0.0}_{-43.2}$ & 1.33$^{+0.26}_{-0.01}$ & 63 & 0.01$^{+0.01}_{-0.01}$ & 0.72 & -140057.35\\
    J033212.23-274620.9 & 2486.3$^{+6.3}_{-6.7}$ & 71.9$^{+0.1}_{-6.8}$ & 199.6$^{+0.4}_{-42.2}$ & 1.45$^{+0.29}_{-0.01}$ & 68 & 0.01$^{+0.01}_{-0.01}$ & 0.36 & -5567.86\\
    J033215.32-275037.6 & 3491.2$^{+19.6}_{-20.2}$ & 71.8$^{+0.2}_{-6.8}$ & 50.1$^{+43.4}_{-0.1}$ & 2.05$^{+0.40}_{-0.02}$ & 96 & 0.01$^{+0.01}_{-0.01}$ & 0.42 & -1261.17\\
    J033222.17-274811.6 & 2799.0$^{+20.3}_{-21.8}$ & 45.4$^{+9.9}_{-7.2}$ & 50.1$^{+42.1}_{-0.1}$ & 3.49$^{+0.89}_{-0.88}$ & 289 & 0.03$^{+0.01}_{-0.01}$ & 0.49 & -91.65\\
    J033222.56-274815.0 & 4147.3$^{+24.7}_{-26.2}$ & 72.0$^{+0.0}_{-6.9}$ & 50.7$^{+42.5}_{-0.7}$ & 2.42$^{+0.49}_{-0.01}$ & 113 & 0.02$^{+0.01}_{-0.01}$ & 0.58 & -69.91\\
    J033235.18-275215.7 & 109.1$^{+4.3}_{-14.4}$ & 57.0$^{+6.2}_{-10.8}$ & 198.8$^{+1.2}_{-90.5}$ & 0.10$^{+0.04}_{-0.02}$ & 6 & 0.00$^{+0.01}_{-0.01}$ & 0.02 & -328.50\\
    J033238.01-274401.2 & 1069.4$^{+19.2}_{-17.5}$ & 37.1$^{+6.6}_{-9.5}$ & 199.9$^{+0.1}_{-43.0}$ & 1.72$^{+0.57}_{-0.32}$ & 167 & 0.02$^{+0.01}_{-0.01}$ & 0.21 & -112.10\\
    J033244.02-274635.9 & 3419.4$^{+38.9}_{-37.5}$ & 45.2$^{+9.9}_{-6.8}$ & 101.1$^{+40.9}_{-41.8}$ & 4.29$^{+1.03}_{-1.07}$ & 357 & 0.04$^{+0.01}_{-0.01}$ & 0.57 & -101.89\\
    J033246.83-275120.9 & 917.1$^{+6.7}_{-7.4}$ & 71.8$^{+0.2}_{-6.7}$ & 199.9$^{+0.1}_{-43.3}$ & 0.54$^{+0.10}_{-0.01}$ & 25 & 0.01$^{+0.01}_{-0.01}$ & 0.22 & -4061.23\\
    J033302.94-275146.9 & 14138.5$^{+225.5}_{-262.1}$ & 37.0$^{+6.6}_{-9.3}$ & 99.6$^{+42.0}_{-41.6}$ & 22.80$^{+7.40}_{-4.20}$ & 2223 & 0.23$^{+0.08}_{-0.05}$ & 2.56 & -65.28\\
    J095905.05+022156.4 & 9410.8$^{+175.9}_{-193.1}$ & 45.1$^{+9.9}_{-6.6}$ & 150.7$^{+41.5}_{-42.4}$ & 11.80$^{+2.80}_{-3.00}$ & 987 & 0.13$^{+0.03}_{-0.04}$ & 1.01 & -4.61\\
    J095929.70+021706.4 & 5070.1$^{+51.2}_{-211.0}$ & 72.0$^{+0.1}_{-13.7}$ & 150.4$^{+45.5}_{-47.6}$ & 2.96$^{+1.19}_{-0.02}$ & 139 & 0.03$^{+0.01}_{-0.01}$ & 0.44 & -2.47\\
    J095945.15+023021.1 & 2282.0$^{+10.1}_{-9.9}$ & 71.9$^{+0.1}_{-6.7}$ & 100.1$^{+41.6}_{-41.4}$ & 1.34$^{+0.26}_{-0.01}$ & 63 & 0.01$^{+0.01}_{-0.01}$ & 0.18 & -706.44\\
    J095953.85+021853.6 & 1705.2$^{+11.1}_{-10.2}$ & 64.2$^{+6.6}_{-6.1}$ & 199.8$^{+0.2}_{-42.7}$ & 1.22$^{+0.23}_{-0.19}$ & 69 & 0.01$^{+0.01}_{-0.01}$ & 0.15 & -297.44\\
    J095959.96+020633.1 & 691.2$^{+5.4}_{-5.3}$ & 45.2$^{+9.9}_{-7.1}$ & 199.8$^{+0.2}_{-42.7}$ & 0.87$^{+0.22}_{-0.22}$ & 72 & 0.01$^{+0.01}_{-0.01}$ & 0.08 & -173.25\\
    J100003.73+020206.4 & 1049.4$^{+31.6}_{-80.1}$ & 71.9$^{+0.1}_{-22.6}$ & 50.3$^{+42.8}_{-0.3}$ & 0.62$^{+0.50}_{-0.02}$ & 29 & 0.00$^{+0.01}_{-0.01}$ & 0.07 & -21.81\\
    J100006.11+015239.2 & 3076.4$^{+31.8}_{-82.1}$ & 71.8$^{+0.2}_{-13.1}$ & 199.5$^{+0.5}_{-45.4}$ & 1.81$^{+0.72}_{-0.03}$ & 85 & 0.02$^{+0.01}_{-0.01}$ & 0.24 & -11.02\\
    J100006.55+023259.3 & 2005.4$^{+13.9}_{-15.9}$ & 72.0$^{+0.1}_{-6.8}$ & 199.9$^{+0.1}_{-43.4}$ & 1.17$^{+0.23}_{-0.01}$ & 55 & 0.01$^{+0.01}_{-0.01}$ & 0.15 & -4308.82\\
    J100033.61+014902.0 & 2553.0$^{+30.1}_{-26.7}$ & 71.9$^{+0.1}_{-6.8}$ & 200.0$^{+0.1}_{-42.9}$ & 1.49$^{+0.29}_{-0.02}$ & 70 & 0.02$^{+0.01}_{-0.01}$ & 0.19 & -327.48\\
    J100055.34+023441.1 & 20213.0$^{+53.9}_{-59.2}$ & 36.8$^{+6.9}_{-9.0}$ & 199.7$^{+0.3}_{-41.9}$ & 32.90$^{+10.20}_{-6.20}$ & 3210 & 0.38$^{+0.12}_{-0.08}$ & 1.52 & -10787.79\\
    J100107.46+015718.1 & 12755.5$^{+126.0}_{-141.9}$ & 26.1$^{+9.2}_{-8.5}$ & 199.6$^{+0.4}_{-42.0}$ & 28.40$^{+7.10}_{-6.60}$ & 3459 & 0.33$^{+0.08}_{-0.08}$ & 1.48 & -8393.19\\
    J100116.15+023606.9 & 6042.4$^{+14.7}_{-16.6}$ & 45.0$^{+10.2}_{-6.7}$ & 199.9$^{+0.1}_{-42.6}$ & 7.62$^{+1.80}_{-1.97}$ & 637 & 0.09$^{+0.02}_{-0.02}$ & 0.61 & -1990.78\\
    J100139.73+022548.5 & 70.0$^{+0.5}_{-0.5}$ & 71.9$^{+0.1}_{-6.9}$ & 198.9$^{+1.1}_{-42.7}$ & 0.04$^{+0.01}_{-0.01}$ & 2 & 0.00$^{+0.01}_{-0.01}$ & 0.06 & -189839.12\\
    \hline
  \end{tabular}
\end{table*}

\begin{figure*}
\includegraphics[width=\textwidth]{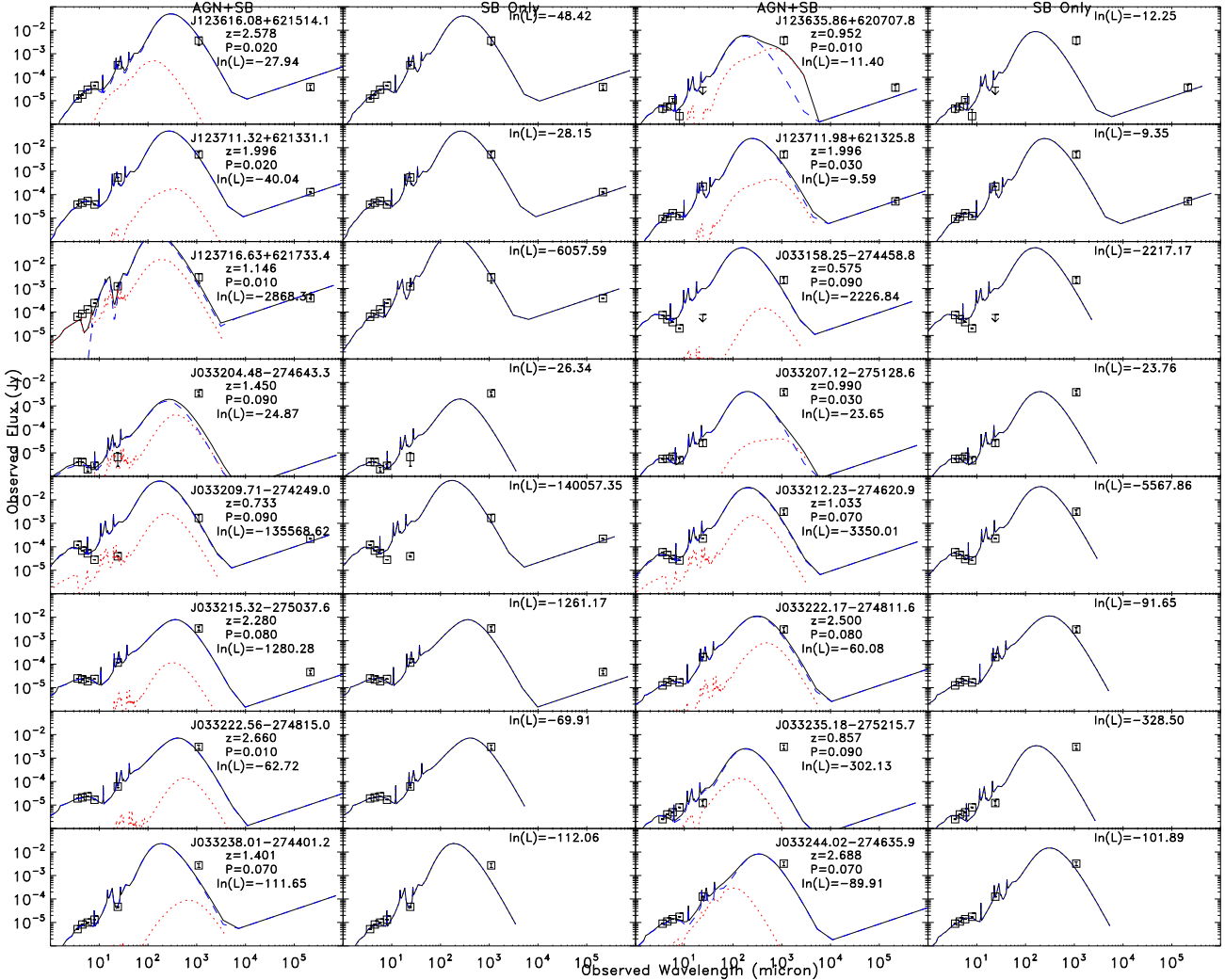}
\caption{Observed frame best-fit AGN+SB and SB-only SEDs to
  AzTEC/X-ray sources.  The plots show the template
  models that lie closest to the best-fit parameters determined from
  the MCMC SED fitting.  The AGN and SB models are given by the
  dotted red and dashed blue lines, respectively, with their linear
  combination shown by the solid black line.  Also shown for
  each source is the redshift used in the SED fitting, favoring the
  spectroscopic redshift where available, and the resulting best-fit
  ln(L) values.  For reference, we have included the
  probability of random association P (Table~\ref{tab:xid}) for each source.}
\label{fig:seds}
\end{figure*}

\begin{figure*}
  \includegraphics[width=\textwidth]{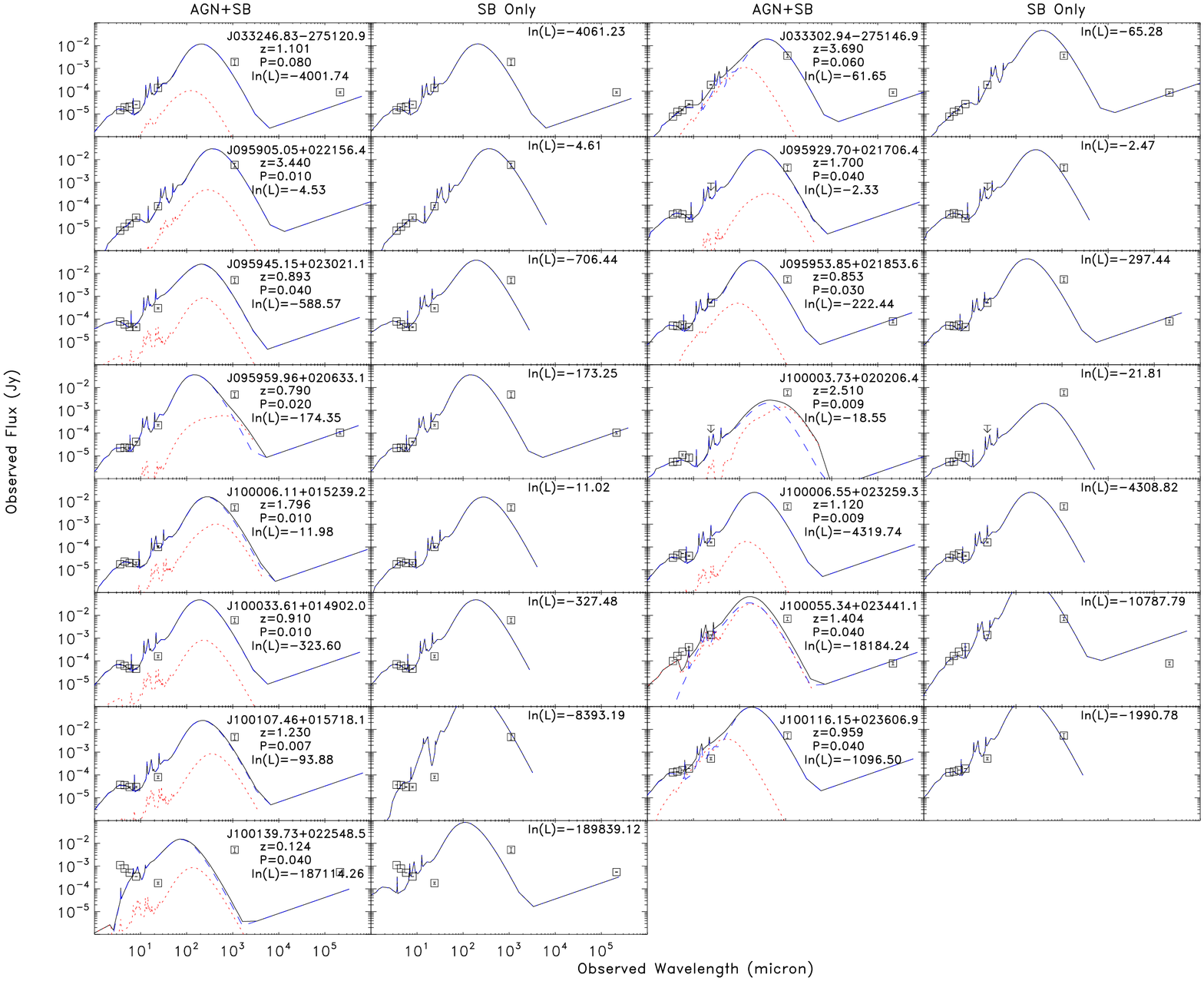}
\contcaption{}
\end{figure*}

\begin{figure*}
\includegraphics[width=0.7\textwidth]{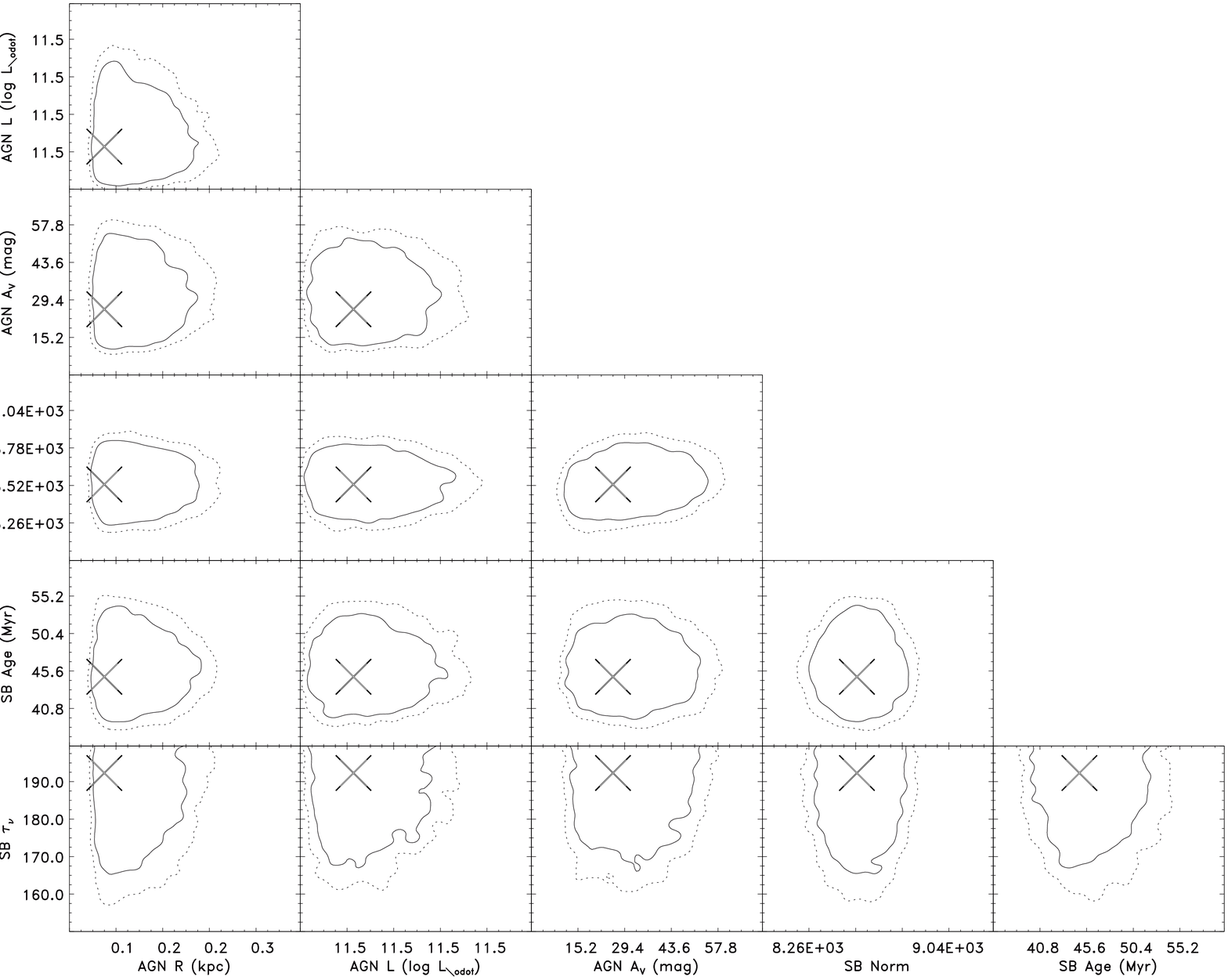}
\includegraphics[width=0.7\textwidth]{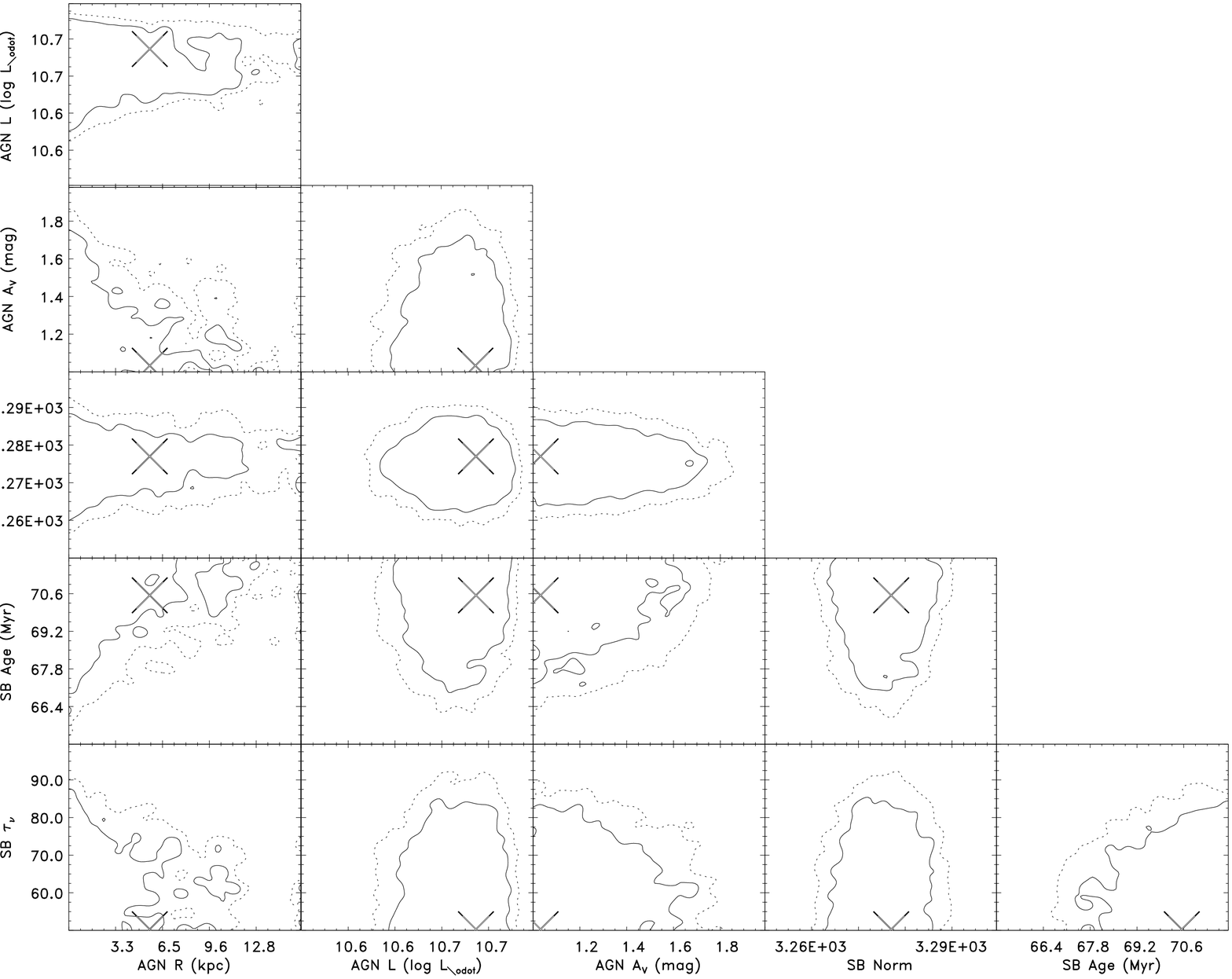}
\caption{Smoothed, marginalized likelihood distribution of accepted
  parameter steps for J123616.08+621514.1 (top) and
  J033215.32-275037.6 (bottom), using the AGN+SB templates. A
  description of each parameter in the AGN and SB templates is given
  in \S~3.2.1.  The location of the maximum likelihood value has been
  marked by the large 'X'.  Contours are drawn at the 68 percent
  (solid) and 90 percent (dashed) confidence levels. The likelihood
  distributions show that while there are no large apparent
  correlations between parameters, the constraints on some
  parameters are rather poor (particularly for AGN R and SB
  $\tau_{\nu}$).  AGN L and SB normalization are the most well
  constrained due to the inclusion of the X-ray luminosity prior.}  
\label{fig:sedcont}
\end{figure*}

\subsubsection{SED Fitting Results}

\begin{figure}
\includegraphics[width=0.5\textwidth]{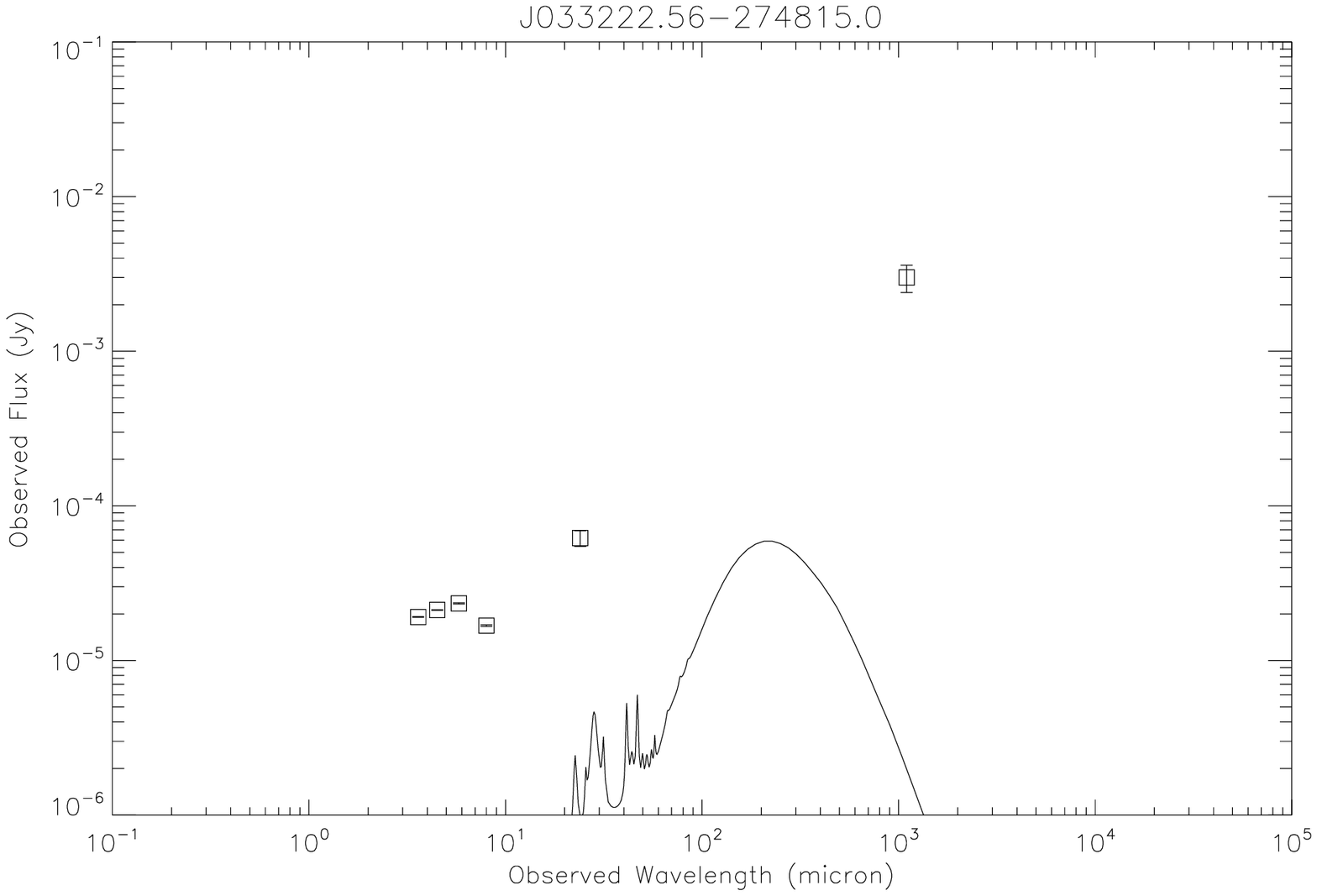}
\includegraphics[width=0.5\textwidth]{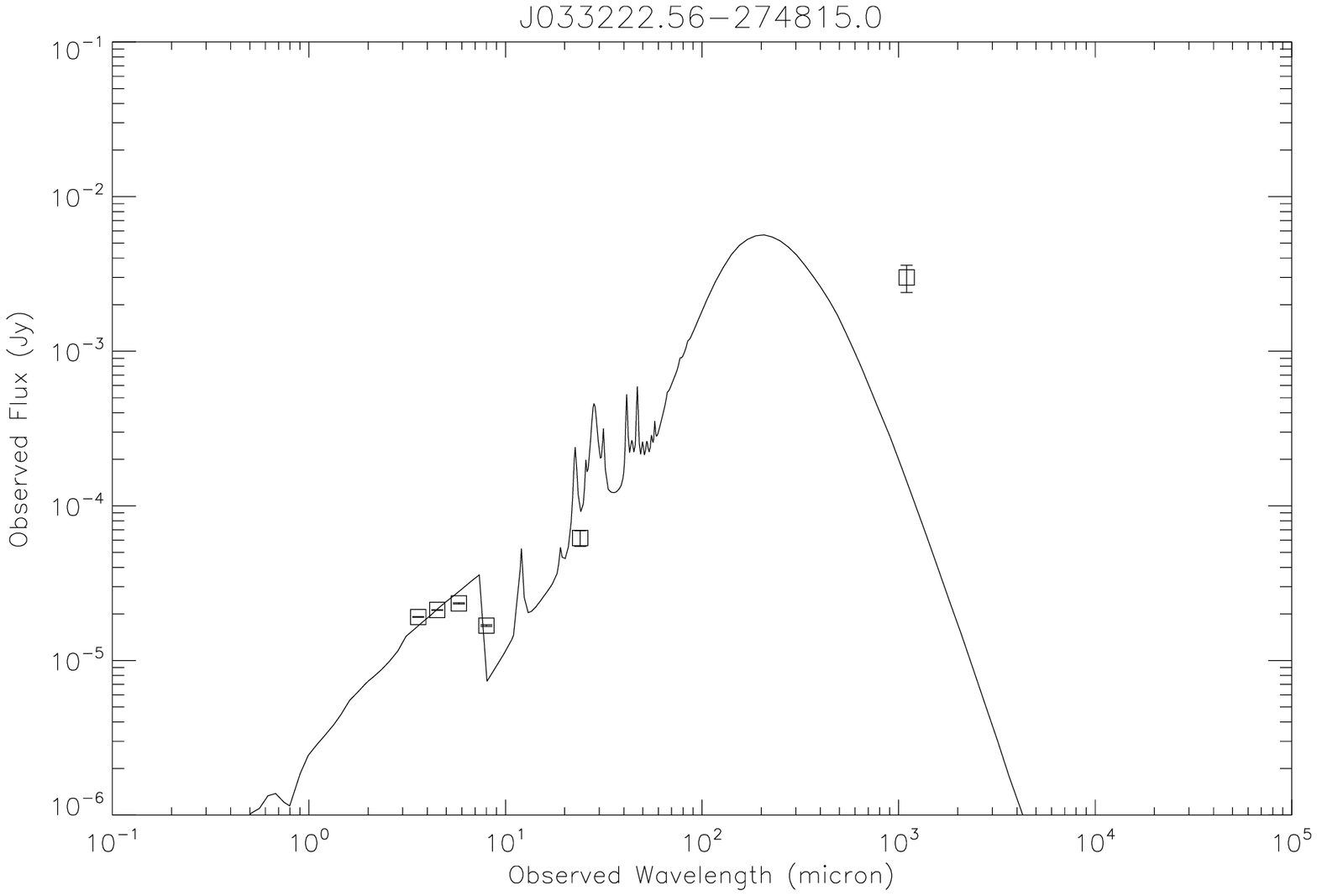}
\caption{AGN only fits for J03222.56-274815.0 using the X-ray priors
  (top) and without (bottom).  As seen in Figure~\ref{fig:seds}, the
  AGN component is unable to contribute any more to the NIR-to-radio
  SED as its vertical scaling, i.e. luminosity, is set by the X-ray
  priors.  Without these constraints, the models are able to account
  for some of the mid-IR emission, still missing the bulk of the
  sub-mm flux, but would predict tremendous X-ray luminosities; here
  the best fit AGN L is $\sim 10^{12.75}$ L$_{\odot}$ which translates
  to an X-ray luminosity of $\sim 5.9\times 10^{44}\rm~ergs~s^{-1}$,
  many times over what is actually observed.}
\label{fig:agn}
\end{figure}

Figure~\ref{fig:seds} shows that our method is able to produce
reasonable fits to the AzTEC/X-ray sources; the source
J033234.78-274815.0 was excluded as it has no discernible
IRAC/MIPS counterpart despite having a spectroscopic redshift (see
Table~\ref{tab:mid}).  While the majority ($\sim$87 percent) of our
sources can be fit using the SB templates alone, they typically
under-predict the 1.1 mm emission, recovering on average $\sim$30-38
percent of the observed flux.  Including the AGN models helps to
slightly increase the model fluxes and are generally required to match
the X-ray luminosity prior but are still unable to match the
mm-wavelength observations; contributing little, if at all, to the
bolometric luminosities and observed 1.1 mm flux.  In some cases
(e.g. J033212.23-274620.9), the AGN and SB templates appear very
similar in the final fit. This likely results from the similarities in
the dust treatment and radiative transfer in the templates as noted by
\cite{siebenmorgen05} for effectively identical template parameters
(i.e. dust content, optical depth and intrinsic luminosity; see also
their figure 4).  In many cases, the fit values for AGN R and
A$_{\rm{V}}$ are rather poor and show a large range in acceptable
values.  This effect stems from our use of the X-ray luminosity prior
which effectively sets the AGN bolometric luminosity, preventing any
additional AGN contribution to the bolometric SED and thus leading to
unconstrained R and A$_{\rm{V}}$ (see also Figure~\ref{fig:agn}).  The
fact that A$_{\rm{V}}$ had such poor constraints prompted us to
include the X-ray column density to avoid over-estimating the dust
content.  By virtue of our MCMC technique, we may readily identify any
degeneracies between the AGN and SB template sets; however,
Figure~\ref{fig:sedcont} shows that there are no large
parameter-parameter degeneracies, although some parameters are not 
very well constrained.  This is particularly the case for AGN R and
A$_{\rm{V}}$ as mentioned above.  

Due to the under-prediction of the 1.1 mm flux, we are biased to under
estimate the total IR luminosity such that the values reported in
Table~\ref{tab:sedas} are more likely to be lower limits.  Correcting
for the under-prediction, we can expect the luminosities to be
$\sim$2-3 times higher.  As a result, the FIR-derived SFRs and the
associated SFR-derived X-ray luminosities for J033215.32-275037.6 and
J033207.12-275128.6 are more likely to be in line with their
observed X-ray luminosities. The remaining starburst-candidate X-ray
sources (J033158.25-274458.8 and J100139.73+022548.5) are still
ambiguous as their X-ray derived SFRs are $\sim$5-8$\times$ higher
than their IR counterparts; however, the poor fits to these
sources prevents an accurate measurement of their FIR luminosity and
SFR, hindering our interpretation. Nevertheless, it remains plausible
that at least $\sim$6 percent of our X-ray-detected SMGs are
starburst-dominated in both the IR and X-ray with little (if any)
emission due to an AGN. 

It is possible to account for the missing 1.1 mm flux if we relax the
constraints on many of the fit parameters.  For instance, the SB
models can provide a better fit if we relax the redshift
prior. Similarly, if the X-ray luminosity constraint is removed then
the AGN templates can account for the remaining 1.1 mm flux with
significantly more dust (as set by A$_{\rm{V}}$ and R).  These fits,
however, are completely unphysical either due to inaccurate redshifts
($\Delta z \gtrsim$0.5) or X-ray luminosity (unconstrained AGN
templates predict orders of magnitude higher X-ray
luminosities, see also Figure~\ref{fig:agn}). Instead, these
additional fits suggest that an additional, possibly extended, dust
distribution may be required. Similar modifications have been
suggested for other SED templates in order to provide complete fits to
other SMGs and millimetre-detected QSOs
\citep[e.g.][]{pope08,martinez09,rowan10}. Unlike \cite{rowan10},
however, we find that a diffuse 'cirrus' component as described by
\cite{efstathiou03} is not sufficient for the additional dust
distribution and does not improve the quality of our fits.    

\section{Discussion}

Across a $\sim$1.2 square degree area of the sky, we have analyzed the
X-ray spectral and NIR-to-radio SED properties of 45 X-ray-detected
AzTEC sources for evidence of AGN and starburst activity.  Our full
sample is limited by the number of available redshifts, leaving a
subset (32/45) of sources.  Within GOODS-N and GOODS-S, this subset of
AzTEC/X-ray sources typically have high levels of dust obscuration
(N$_{\rm{H}}\gtrsim10^{23}\rm~cm^{-2}$) and are generally associated
with AGN activity, while their NIR-to-radio SEDs imply that the IR and
bolometric output are almost completely dominated by star formation.
Though we do go deeper in the 4Ms GOODS-S field and find fainter
potential X-ray counterparts, we do not find any evidence for
significantly higher amounts of dust obscuration compared to the 2Ms
GOODS-N and initial 2Ms GOODS-S. Considering the relative
uncertainties in the L$_{\rm{X}}$-SFR relation and under-prediction of
the 1.1 mm flux for many of our models, a small portion ($\sim$6-13
percent) of our X-ray-identified SMGs are likely to be completely
dominated by starburst emission in both the X-ray and NIR-to-radio
with the remaining majority powered almost exclusively by an
AGN in the X-ray and starburst in the NIR-to-radio (see \S~3.1.3 and
\S~3.2.2).  Here, we explore the implications of our X-ray modeling
and SED fitting in the context of emission at 1.1 mm and previous
(sub-)mm/X-ray studies. 

\subsection{Origin of 1.1 mm Emission}

As stated in \S~3.2, (sub-)millimetre emission from our AzTEC/X-ray
sources results from dust heated to T$\sim$30K.  Based on our
SED fitting (\S~3.2.2), the observed NIR-to-mm luminosity is generally
dominated by the starburst with little contribution from an AGN (see
Figure~\ref{fig:seds}). These fits predict dust temperatures on the
order of $\sim$30-40K yet generally under-predict the observed 1.1 mm
flux by $\gtrsim$50 percent, suggesting that a cooler, extended dust
component is present.  Further evidence for an additional dust
component has been seen in previous SMG studies
\citep[e.g.,][]{chapman05,pope06} and by \cite{rowan10} in
\textit{Herschel} SPIRE sources, although they suggest that it can be
accounted for with the cirrus templates of \cite{efstathiou03} which
we are unable verify. Fitting of the IR dust peak, for which
\textit{Herschel} data are optimized, will provide more accurate
estimates of the dust temperature and will aid in reducing parameter
uncertainty, improving the bolometric luminosity estimates and
providing further insight into the nature of the missing dust.  

Given the evidence so far for an additional dust component, one
must wonder where the dust resides.  The dust could simply reside in
an extended disk if the starbursting region remains localized to the
central $\sim$1 kpc.  Alternatively, the dust may reside in the halo
of the SMG, pushed out through radiation- or momentum-driven outflows
resulting from the starburst region(s) and/or the central AGN
\citep[e.g.,][]{oppen06,zu11}.  A typical GMC in a $z=2$ starburst
galaxy can reach velocities of $\sim$300 km s$^{-1}$ \citep{murray11},
which will spread its gas and dust as far as $\sim$15 kpc from the
galaxy center during a $\sim$50 Myr starburst active phase.  Similar
outflows reaching $\sim$1000 km s$^{-1}$ have been observed in local
ULIRGS and have been shown to account for as much as 20 percent of the
total molecular gas mass, on the order of 10$^9$ M$_{\odot}$, which
are easily produced through starbursts with SFR$\gtrsim
100\rm~M_{\odot}~yr^{-1}$ \citep{chung09,chung11}.  The spatial scales
predicted for these outflow regions are consistent with the radii
predicted by the AGN templates and high resolution imaging of SMGs
using the IRAM Plateau de Bure interferometer (PdBI) and Submillimetre
Array (SMA) \citep[e.g.,][]{tacconi06,tacconi08,younger08}, which show
typical size scales of $\sim$2-8 kpc.  The molecular gas will not
survive long due to lack of self-shielding, which would allow the dust
to inhabit a larger volume than that traced by traditional molecular
gas measurements. Though the majority of the mass will still be
contained within the central region, the extended dust will quickly
cool to the background temperature and will likely produce a
temperature gradient as distance from the central region increases
\citep[see, for example, fig. 5 of][]{younger08}, which may contribute
significantly to the (sub-)mm emission.  This scenario agrees with the
recent EVLA observations of \cite{ivison10,ivison11} SMGs and of
background quasars \citep{menard10}.

A possible alternative to the extended cold dust model is that
the missing 1.1 mm flux results from either false detections or source
blending.  We may readily remove false detections as the false
association rate for the X-ray-identified AzTEC sources is $\sim$5-6
percent (\S~2.3.1) whereas the majority of sources under-predict
the 1.1 mm flux.  Similarly, previous sub-mm studies suggest blending
can occur in $\sim$20-25 percent of sub-mm detected sources
\citep[see][and references therein]{scott12} so that while blending is
likely to occur, it is unlikely to the primary cause for the flux
discrepency.  Unfortunately, it is not possible to
de-blend sources using current \textit{Spitzer} MIPS,
\textit{Herschel} PACS or VLA radio data without \textit{a priori}
knowledge of the intrinsic sources.  Only through high resolution
imaging and kinematics with ALMA, LMT and future (sub-)mm telescopes
may we be able to de-blend potential offenders and/or make direct
confirmation of an outflow-produced, extended cold dust distribution.

The question still remaining is how the AGN emission, as indicated by
the high X-ray detection rate and the X-ray spectra, relates to the
sub-mm observations.  While AGN models are favored in the X-ray
spectral fitting, the sub-mm emission is, in fact, unlikely
to result solely from an AGN.  As shown in Figure~\ref{fig:agn}, the
X-ray priors prevent any significant contribution from the AGN
templates.  Even when relaxed, the AGN models still show poor fits to
the mid-IR, relative to their observed fluxes and uncertainties, and
sub-mm data, never mind the unphysical X-ray luminosity predicted.  In
a merger-driven formation scenario \citep[e.g.][]{nara10},  gas from
the colliding systems gives rise to an increase in star formation,
resulting in a sub-mm-bright phase. Shortly after the sub-mm-bright
phase and final coalescence, the central black hole may undergo the
bulk of its growth, producing an AGN which may then aid in shutting
off the star formation through feedback, leaving the final system as a
quasar or dusty AGN-powered ULIRG.  Given the high X-ray column
densities we derived for our AzTEC/X-ray sample, it is likely that
these sources represent the early growth phase of the AGN.  Combined
with the starburst-dominated NIR-to-radio SEDs and expected short
timescale of the sub-mm-bright phase ($<50$ Myr), the X-ray-identified
sources may be SMGs caught in their transitionary period between peak
star formation and peak black hole growth.  This transition scenario
is consistent with the fact that the average 1.1 mm fluxes and 2-10
keV count rates of the X-ray-identified sources are both below the
average of the overall SMG and X-ray sample. The remaining X-ray
undetected SMGs could result from a starburst triggered during the
first passing of merging systems or rapid, short-lived mergers similar
to those found by \cite{chapman09}. However, we can not rule out the
possibility that the X-ray-dim SMGs could result from a moderately
continuous gas in-fall \citep[see, for example, ][]{dave10} or very
young starbursts with Compton-thick AGNs (e.g., A05a,b).  The
X-ray-detected SMGs are unlikely to be produced by such continuous
in-fall given the starburst timescales from our SED modeling;
accretion-driven models predict that the sub-mm bright phase may last
for $\sim$0.1-1 Gyr \citep{fardal01}. One other possibility given the
expected high SFRs for SMGs is that the central AGN are likely
time-variable \citep[see][and references therein]{alex12}. AGN can
switch between being 'on' or 'off' on timescales of $\lesssim$1 yr and
cause large variations in their observed luminosities and absorbing
column densities, which will affect the probability of detecting an
AGN associated with an SMG.  It is unknown how this AGN
time-variability scenario will influence the SED of SMGs though we
expect any contribution to be small given the already low AGN
contribution rate.  Further evolutionary simulations and observations
aimed at spatially resolving SMGs will provide the tools necessary to
classify SMGs under the appropriate formation and evolutionary scenario.   

\subsection{Comparison with Previous Studies}

\begin{figure}
\includegraphics[width=0.5\textwidth]{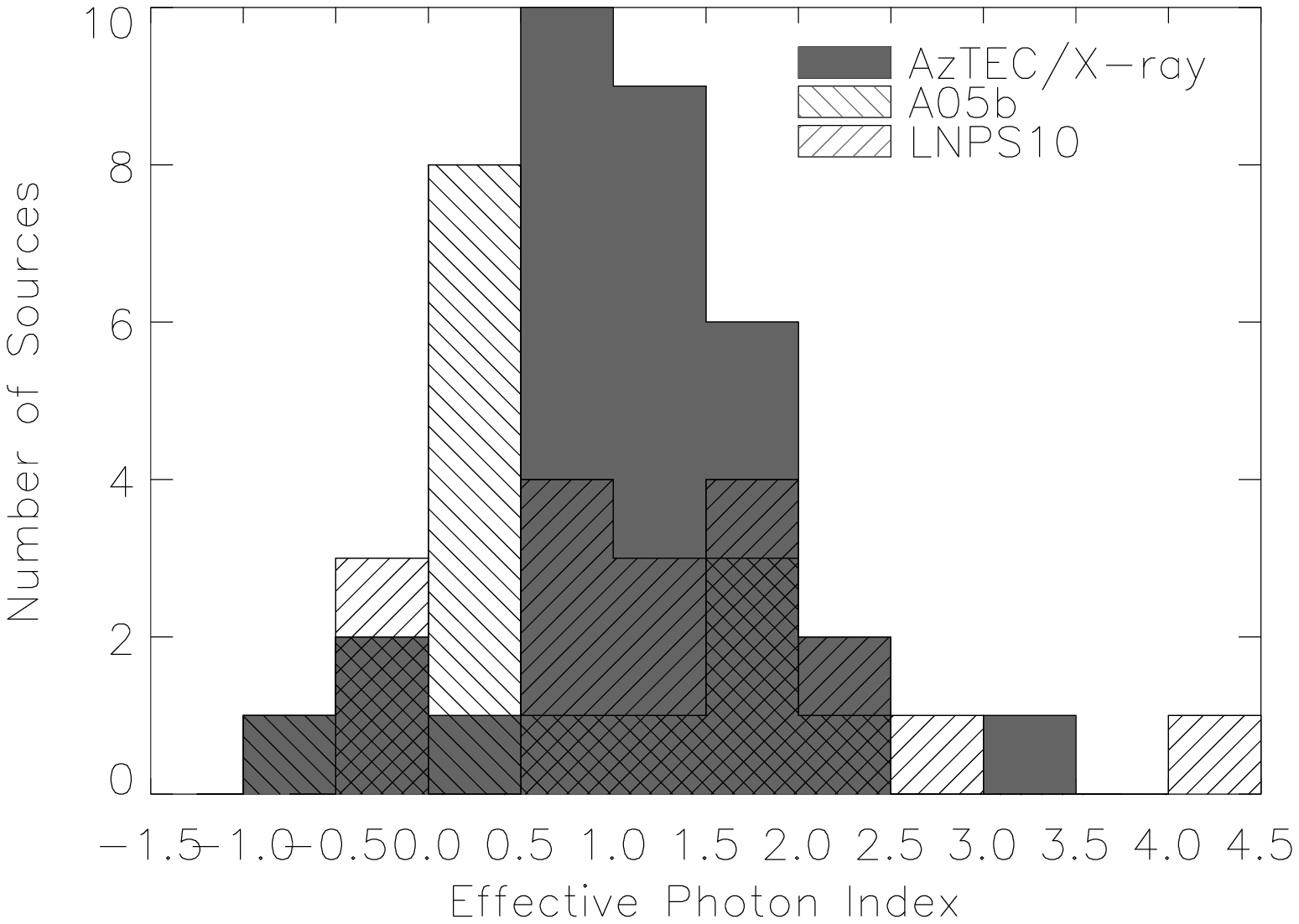}
\includegraphics[width=0.5\textwidth]{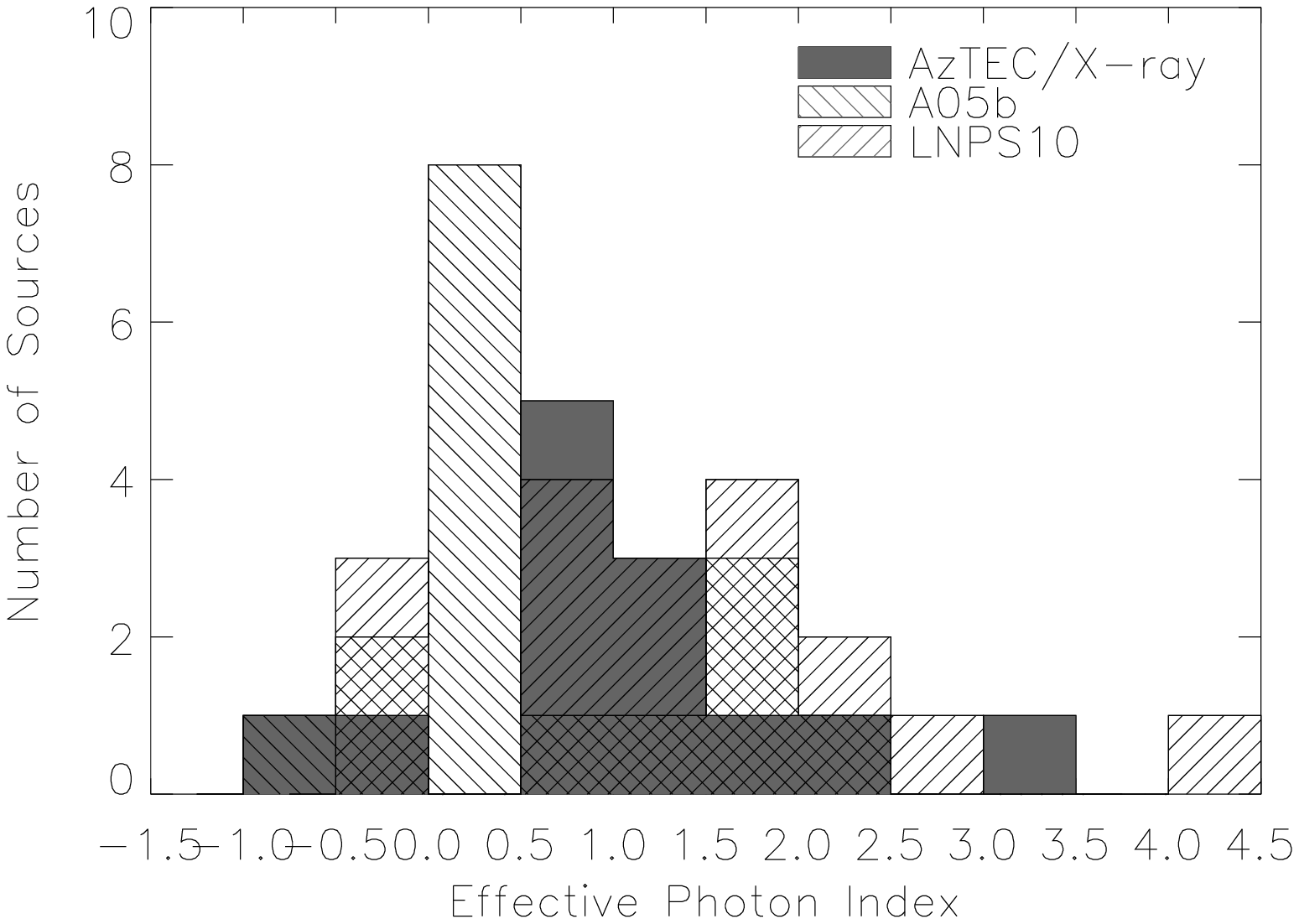}
\caption{Histogram of effective power-law indexes
  ($\Gamma_{\rm{Eff}}$) for the AzTEC/X-ray sources, given by the
  filled histogram, for all sources (top) and only
  those with radio counterparts (bottom).  For comparison, we also
  include the $\Gamma_{\rm{Eff}}$ distributions from A05b
  (back-hashed region) and LNPS10 (forward-hashed region).}
\label{fig:effgamma}
\end{figure}

\begin{figure}
\includegraphics[width=0.5\textwidth]{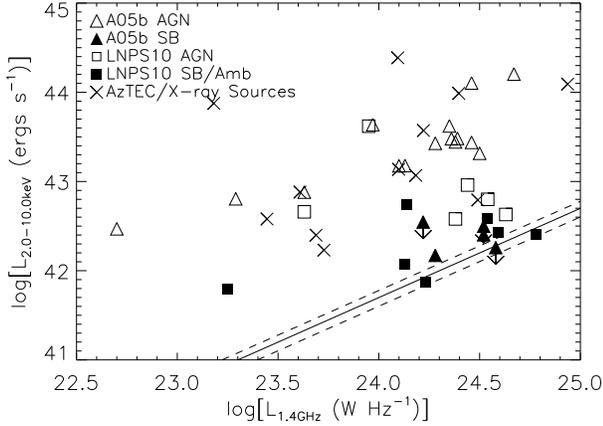}
\caption{Reproduction of fig. 2b from A05b including the A05b, LNPS10
  and AzTEC/X-ray samples.  The fluxes of
  A05b have been converted to 2.0-10.0 keV luminosities
  assuming a photon index of $\Gamma=1.8$ and eqn. 1 of
  \citet{alex03}.  Radio luminosities are calculated from the radio
  fluxes in Table~\ref{tab:mid} and eqn 2. of \citet{alex03}.  Also
  plotted is the P04 SFR-X-ray luminosity relation (solid line), using
  the SFR-radio relation of \citet{condon92} to convert SFR to 1.4 GHz
  luminosity, with a 20 percent statistical error given by the
  dotted lines.  Some of the A05b starburst sources only
  have 3$\sigma$ upper limits for their X-ray luminosities and are
  shown with arrows indicating such.} 
\label{fig:a052b}
\end{figure}

\begin{figure}
\includegraphics[width=0.5\textwidth]{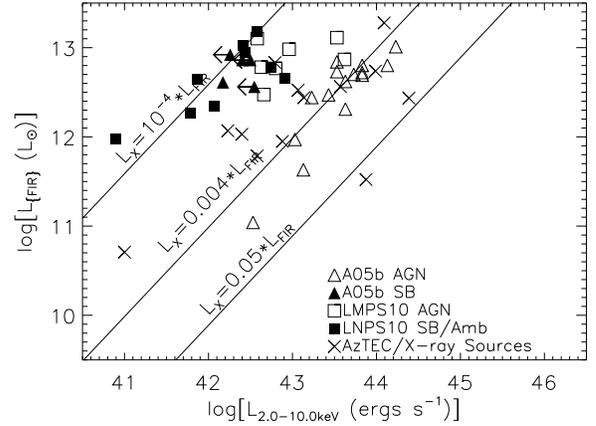}
\caption{Reproduction of fig. 8 from A05b including the A05b, LNPS10
  and AzTEC/X-ray samples. The X-ray fluxes of A05b are converted to
  2.0-10.0 keV using $\Gamma=1.8$.  FIR luminosities are derived from
  the radio luminosities of Fig.~\ref{fig:a052b} using a radio to FIR
  correlation of q=2.35 \citep{helou85}.  The over-plotted lines
  represent ratios of constant X-ray versus FIR luminosity for the
  A05b starburst ($\frac{\rm{L_X}}{\rm{L_{FIR}}}=10^{-4}$) and AGN
  ($\frac{\rm{L_X}}{\rm{L_{FIR}}}$=0.004) sources, and the average
  luminosity ratio for quasars studied by \citet{elvis94}
  ($\frac{\rm{L_X}}{\rm{L_{FIR}}}$=0.05).}
\label{fig:a058}
\end{figure}

For our AzTEC/X-ray sample, the AGN detection rate is $\sim$14 percent
between all three fields.  However, the shallow X-ray depth of COSMOS,
potentially compounded by our more stringent detection criteria,
prevents confirmation of the most heavily obscured AGNs which may
contribute significantly to the sub-mm emission
\citep{lutz10,hill11}. Excluding COSMOS, the AGN detection rate
increases to $\sim 28$ percent, consistent with previous X-ray/SMG 
studies ($>38^{+12}_{-10}$ percent, 29$\pm$8 percent and $<26\pm 9$
for A05a,b, LNPS10 and GRC11, respectively) while avoiding potential
biases due to prior counterpart identification and achieving better
source statistics via larger sky coverage.  Similar to LNPS10, we also
find evidence that $\sim$6-13 percent of our X-ray sources are
potentially HMXBs associated with high star formation
rates.  However, many of the starburst powered SCUBA-detected sources
of LNPS10, and by extension A05b, are missing from our sample.  While
the our X-ray data for GOODS-N is essentially the same as that used in
A05b and LNPS10, it is not surprising for differences to exist between
the AzTEC and SCUBA catalogs.  \cite{chapin09} suggests that such a
discrepancy results from instrument and measurement calibration
uncertainty as well as intrinsic spread in host properties (namely
dust temperature and emissivity).  In fact, for a SCUBA source to be
detected by AzTEC at $>3.5\sigma$ in GOODS-N (where the AzTEC rms is
$\sim$1.3 mJy/beam, see \S~2.1), its effective 850$\mu$m flux would
need to be $\gtrsim8.19$ mJy, higher than the typical 850$\mu$m flux
for sources in LNPS10.  This estimate assumes an R=S$_{850}$/S$_{1.1}$
value of 1.8 \citep{chapin09} and that 'flux boosting' \citep[see, for
  example,][]{austermann10,scott10} effects the 850$\mu$m and 1.1mm
observations equally.  Completeness of the (sub-)mm observations may
also contribute to this discrepancy; at $\sim$4mJy, the AzTEC map is
$\sim$60 percent complete \citep[][]{perera08}.  Of course, there is
always the issue of false identifications and mismatching of sources
as well as prior counterpart bias (see LNPS10) which, while the
expected number of such occurrences are small (see \S~2), may still
lead to a decrease in the number of starburst-dominated X-ray sources
in our sample. 

In Figure~\ref{fig:effgamma}, we show the range of effective photon
indexes $\Gamma_{\rm{Eff}}$ for our AzTEC/X-ray sample (see \S~3.1.1,
Table~\ref{tab:xspec}) in relation to the samples of A05b and LNPS10.
Using the Mann-Whitney (MW) U-test, we find that the probability that
our AzTEC/X-ray sources are consistent with being drawn from the
samples of A05b and LNPS10 are 0.02 and 0.14, respectively. If we
limit our sample to AzTEC/X-ray sources with radio detections then the
MW probabilities become 0.07 and 0.17 for the A05b and LNPS10 samples,
respectively.  Since the errors on $\Gamma_{\rm{Eff}}$ are known, we
further estimate the intrinsic mean and variance of the samples by
constructing 1000 Monte Carlo realizations of the $\Gamma_{\rm{Eff}}$
distributions. The resulting intrinsic mean value of $\Gamma_{\rm{Eff}}$ for
the AzTEC/X-ray, A05b and LNPS10 samples are 1.14$\pm$0.09
(1$\sigma$), 0.60$\pm$0.10 and 1.44$\pm$0.16, respectively; including
only the radio-detected AzTEC/X-ray sources results in an intrinsic
mean of 1.05$\pm$0.08.  These results imply a strong statistical
difference between the AzTEC/X-ray and A05b samples (at
$\gtrsim$3$\sigma$), while the AzTEC/X-ray and LNPS10 samples have
consistent means values of $\Gamma_{\rm{Eff}}$.  

Despite the differences in $\Gamma_{\rm{Eff}}$, the methods of
analysis in A05b produce results consistent with our study.
For further comparison, we reproduce figures 2b and 8 of A05b, which
show the L$_{\rm{2.0-10.0 keV}}$ versus L$_{\rm{1.4 GHz}}$
(Figure~\ref{fig:a052b}) and L$_{\rm{2.0-10.0 keV}}$ versus
L$_{\rm{FIR}}$ (Figure~\ref{fig:a058}) relations for the A05b and LNPS10
starburst and AGN systems, including our AzTEC/X-ray sources that have
radio counterparts. In reproducing the A05b figures, we have converted
the A05b 0.5-8.0 keV fluxes to 2.0-10.0 keV luminosities assuming a photon
index of $\Gamma=1.8$ and eqn. 1 of \cite{alex03}.  Radio and FIR
luminosities have been determined for our sample following the same
procedures as A05b to ensure compatibility.  We caution, however, that
the radio-FIR correlation used to derive the FIR luminosities from the
radio emission \citep[][]{helou85} assumes emission purely from star
formation and could be misleading if the AGN is radio-loud
\citep[e.g.][]{donley05,donley10}.  Figures~\ref{fig:a052b} and
\ref{fig:a058} show that the X-ray emission for the sub-sample of
radio-identified AzTEC/X-ray sources is higher than one would predict
from their radio and/or FIR luminosities if they resulted purely from
star formation, indicating AGN activity. However, the FIR luminosities
are generally higher than expected for typical quasars which suggests
significant contribution from star formation, again consistent with
the results from \S~3.2.  Alternatively, sources could lie above the
\cite{elvis94} quasar relation if they are reflection dominated or
Compton-thick (e.g. FSC 10214+4727 A05b, Arp 220 \citealt{iwasawa05}).
This is not likely to affect our analysis based on the results from
our X-ray spectral modeling (\S~3.1); nevertheless, we can not rule
out the possibility that the faintest X-ray sources may be harboring
highly luminous, Compton thick AGNs, particularly for the
non-X-ray-detected SMGs \citep[e.g.][]{iwasawa05,lutz10,hill11}.   

\subsection{Cross-Correlation of AzTEC/X-ray source populations}

\begin{figure}
\includegraphics[width=0.5\textwidth]{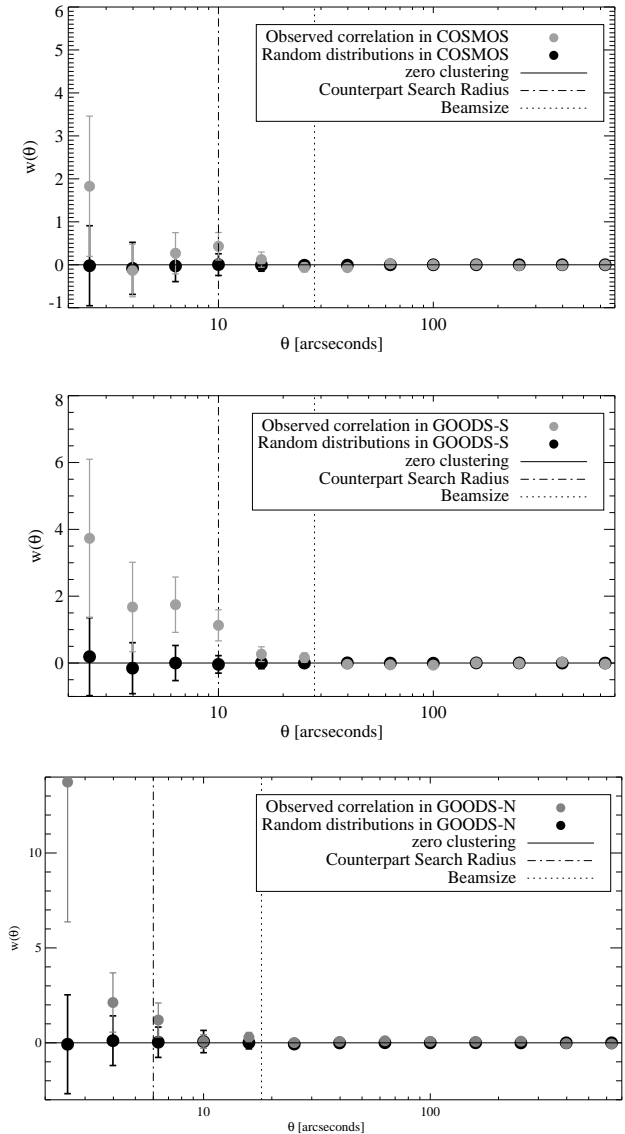}
\caption{The XCF between AzTEC and \textit{Chandra} source
  populations.  Plotted in each panel is the observed XCF and the XCF
  from randomly generated source populations along with the respective
  beam-size and search radius for each field.  Below our adopted search
  radii, the XCF shows significant signal due to detected
  counterparts.  The lack of consistent positive correlation in COSMOS
  results from the shallow X-ray depth and corresponding low source
  density.}
\label{fig:xcf}
\end{figure}

\begin{figure}
\includegraphics[width=0.5\textwidth]{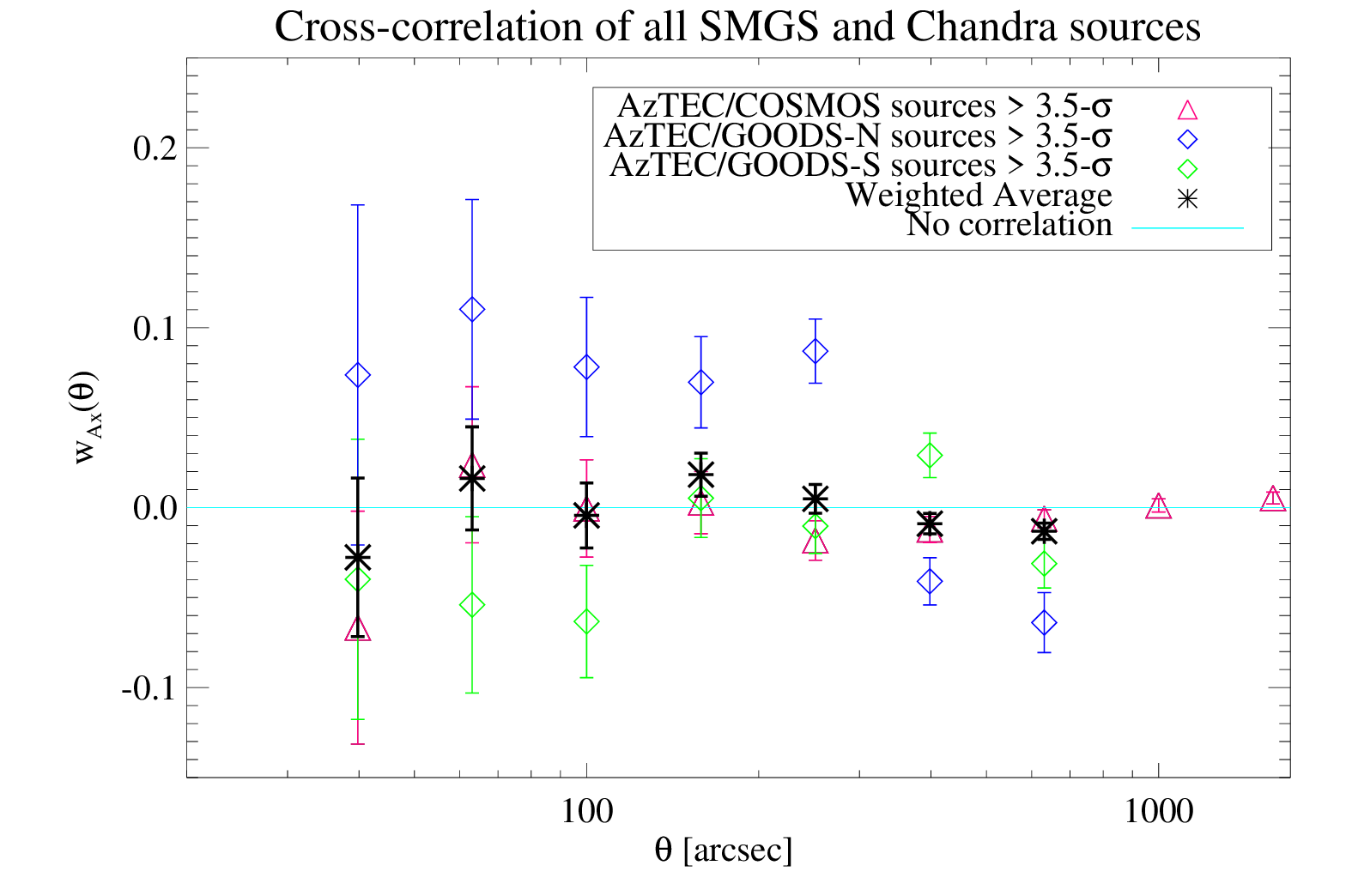}
\includegraphics[width=0.5\textwidth]{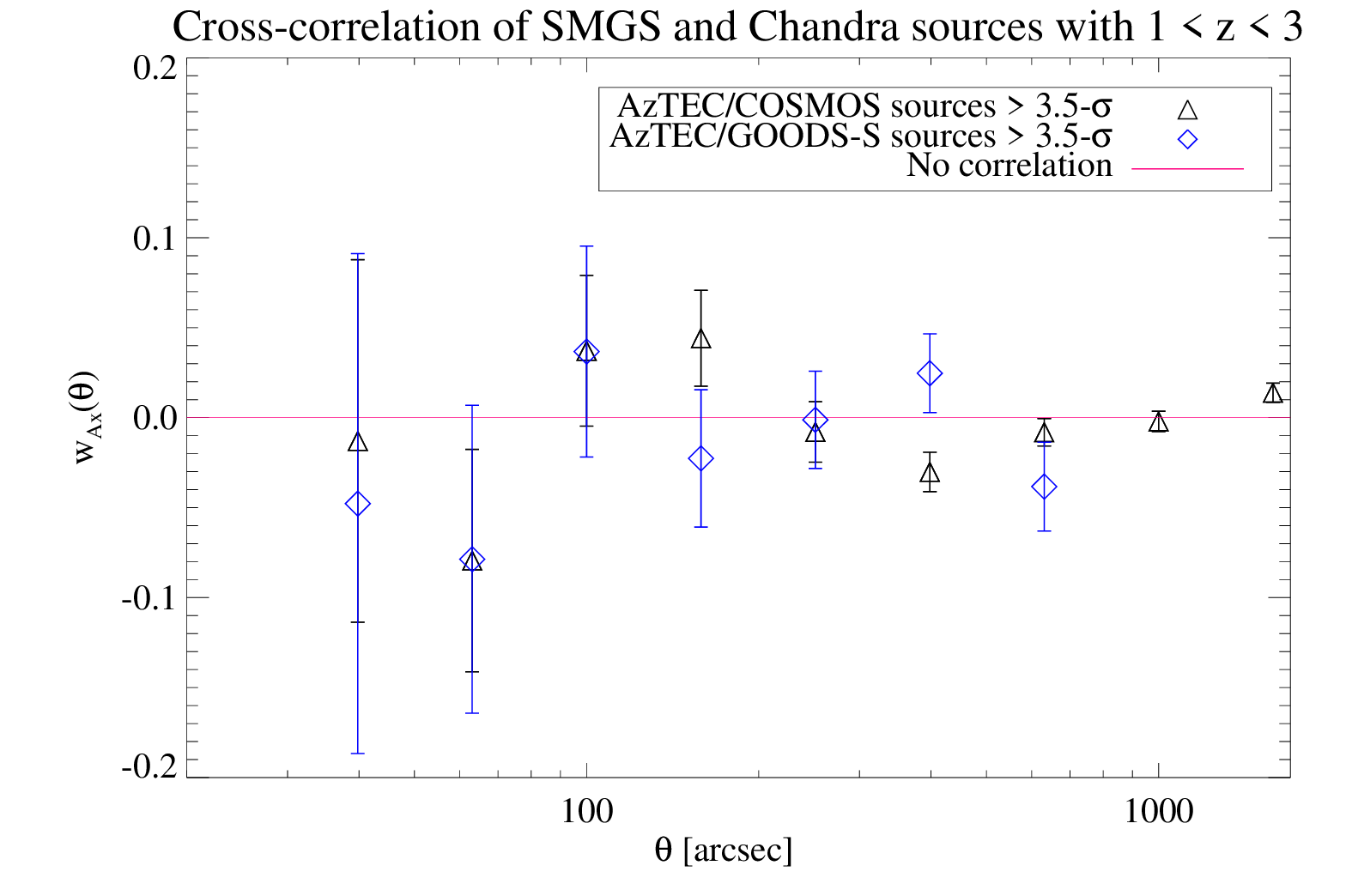}
\caption{The XCF between the AzTEC and \textit{Chandra} source
  populations (shown in Figure~\ref{fig:xcf}) at scales larger than
  the beam-size.  The top panel shows the full sample while the bottom
  shows the XCF in the redshift range 1$<z<$3.  Over the three fields,
  there is no significant correlation between the source populations,
  particularly over the typical redshift range of SMGs.  Due to
  sensitivity variations in the \textit{Chandra} data, the XCF should
  not be heavily weighed at angular separations of $\gtrsim$200.}
\label{fig:xcf_large}
\end{figure}

In addition to examining X-ray-detected SMGs on a source-by-source
basis, a simple cross-correlation analysis of the X-ray and AzTEC
source populations can identify evolutionary patterns between the
populations.  \cite{almaini02,almaini03} were the first to measure the
angular cross-correlation function (XCF) between SMGs and X-ray
sources and found significant correlation at large scales, leading to
the conclusion that the populations both reside in similarly massive
dark matter halos and trace the same large scale structure.
\cite{hill11} later estimated the XCF for LABOCA sources in the ECDFS
and while found similar evidence for small scale clustering,
i.e. residing in same dark matter halos, found no evidence for large
scale clustering, consistent with \cite{borys04}. \cite{roche03}
measured the XCF of extremely red objects (EROs) -- which may be the
signature of massive galaxies that have entered their passive post-AGN
phase in galaxy evolution -- and X-ray sources in the CDFS, and again
find evidence for significant correlation.  Together, these results
may suggest an evolutionary sequence between these three populations,
where starburst dominated SMGs go through an AGN-bright phase before
evolving into passive ellipticals or EROs.

To determine if there is any correlation between the AzTEC and
\textit{Chandra} source populations, we apply the two-point angular
XCF, $w_{AX}(\theta)$, defined as the excess probability of finding
both an AzTEC source in a solid angle $\delta \Omega_{A}$ and an X-ray
source in a solid angle $\delta \Omega_{X}$, with an angular
separation $\theta$ from each other. This excess probability (relative
to an uncorrelated distribution) is given by $\delta P = \rho_{A} \rho_{X}
[1+w_{AX}(\theta)]\delta\Omega_{A}\delta\Omega_{X}$, where $\rho_{A}$ 
and $\rho_{X}$ are the surface densities of AzTEC and X-ray galaxies
on the sky \citep{peebles80}. In practice this can be measured from
galaxy maps by counting the number of SMG/X-ray source pairs, binned
by their angular separation, and comparing to pair counts from
random positions.  Here, we use the cross-correlation
adaptation to the Landy-Szalay estimator \citep{landy93}, which is
given by  
\begin{equation}
w_{AX}(\theta) =
\frac{D_{A}D_{X}(\theta)-D_{A}R_{X}(\theta)-D_{X}R_{A}
  (\theta)+R_{A}R_{X}(\theta)}{R_{A}R_{X}(\theta)}
\end{equation}
where $D_{A}D_{X}$ is the number of SMG/X-ray source pairs,
$D_{A}R_{X}$ and $D_{X}R_{A}$ are the number of pairs found between
each galaxy catalog and randomly generated positions of sources
within each angular separation bin. $R_{A}R_{X}$ is the number of
pairs found between random positions for each galaxy population,
generated from the selection function and sensitivity distribution of
each map. To generate random source distributions for the AzTEC
maps, we follow the methods of \cite{williams2011}. For the
\textit{Chandra} random catalogs, the exposure maps are relatively
uniform (ignoring effects due to CCD gaps and edge overlapping in
COSMOS as they are generally small) such that we may produce the
random catalogs by simply randomly generating positions within the
overlapping coverage region of the \textit{Chandra} and AzTEC maps.
Note, however, that this does not take into account the sensitivity
variations (mostly due to PSF degradation) as a function of off-axis
distance; the XCF may thus be incorrect at scales larger than
$\sim$200$\arcsec$.  The overlapping observations in COSMOS helps to
smooth the telescope response, allowing for a more accurate XCF at
larger scales. 

The resulting XCF for each field, as well as the expected XCF from
completely random distributions, is shown in Figure~\ref{fig:xcf},
where the errors are estimated from a Poissonian distribution given
the number of AzTEC/X-ray pairs in each angular bin.  The expectation
from random distributions is estimated by averaging the XCF of 100
AzTEC and X-ray random distributions described above, which have the
same properties (area and source density) as the observed maps.  In
the case of the random expectation, the errors correspond to the
standard deviation of the XCF from each of the individual random
distributions.  At small scales, there is significant positive
correlation in the observed XCF due to identified counterparts
\citep[see also][]{hill11}; this effect is diluted in COSMOS due to
its shallow X-ray depth and thus low source density compared to either
GOODS field.  However, since the AzTEC source positions are not well
known on scales smaller than the beam-size, we choose to limit our XCF
analysis to the large scale clustering. Figure~\ref{fig:xcf_large}
shows the same XCF combined with their weighted average for scales
larger than 28$\arcsec$, the beam-size of AzTEC on ASTE, though we
caution against heavy interpretation at scales larger than
$\sim$200$\arcsec$ as previously mentioned. 

Across the three fields, we find no evidence for any large scale
correlation signal; any apparent correlation or anti-correlation seen
in individual fields is detected at $\lesssim 1\sigma$ confidence,
consistent with \cite{borys04} and \cite{hill11}. The large area
covered by our sample ($\sim$1.2 square degrees) aids in mitigating
the effects of cosmic variance, which is the likely cause of variation
between fields and may affect the positive correlation signal found in
\cite{almaini02}.  It is possible that the lack of any correlation
signal in our data may be the result of dilution given the wide and
differing redshift distributions of the X-ray and sub-mm
sources. In an attempt to improve the cross-correlation signal, we
have run the same analysis by limiting the X-ray sources to the
redshift range of $1<z<3$ where the X-ray redshift distribution shows
significant overlap with the sub-mm distribution. If there is any
cross-correlation between the two samples, it should be maximized
here. Due to the small number of X-ray sources with available
redshifts in GOODS-N (49 out of the original 397), we excluded this
field from the XCF in the 1$<z<$3 redshift range.  The XCF using this
redshift-limited subset for GOODS-S and COSMOS is statistically
identical to the result we measured using the entire set of X-ray
sources, i.e. no evidence for a correlation.  

The lack of a significant correlation between the X-ray and AzTEC
source populations at large scales may suggest that SMGs and
AGN are not universally related in terms of dark matter halo mass
and large scale structure. However, considering the significant
fraction of AzTEC SMGs that do have plausible X-ray detections here,
it is likely that the SMG phenomenon is not governed by a single
formation and evolution process; rather, the SMG population is a
"mixed bag" of systems -- some undergoing major mergers concurrent
with the build-up of massive black holes \citep[e.g.,][]{nara10} and
others signaling a short-lived phase of intense star-formation in more
normal galaxies \citep[e.g.,][]{chapman09} or even quiescent mass
build-up from gas in-fall \citep[e.g.,][]{dave10} (see also
\S~4.1). Such cases are likely tied to the host's intrinsic properties
which could naturally explain the enhanced sub-mm emission from
bright, obscured AGNs as found by \cite{lutz10} and
\cite{hill11}. However, we caution that limitations in measuring the
correlation between these populations can also give a null result. For
example, the large volume sampled coupled with the lack of redshift
information for the full X-ray and SMG catalogs will necessarily
dilute the {\it projected} correlation strength between the two
populations, even if there is some spatial correlation.  The
shallow X-ray depth of COSMOS will further dilute any correlation
signal by primarily detecting bright AGN that are likely well past
their starburst phase. Observations of SMGs in the near-future with
ALMA and the LMT geared towards measuring their redshifts and
obtaining high-resolution imaging of their dust and gas will greatly
aid in the development and fine tuning of formation and evolution
scenarios for this population.  

\section{Summary}

We have presented a detailed analysis of the X-ray properties of
AzTEC 1.1 mm sources found in the GOODS-N, GOODS-S and COSMOS fields.
Thanks to deep ($\sim$2-4 Ms) \textit{Chandra} observations, we find
X-ray counterparts to $\sim$14 percent of the 1.1 mm sources across
all three fields, increasing to $\sim$28 percent if we exclude COSMOS
due to its shallower X-ray data.  From our modeling of the X-ray
spectra and NIR-to-radio SEDs, we conclude that AzTEC/X-ray sources
are all starburst-dominated in the IR, with SFRs on the order of
$100-1000\rm~M_{\odot} yr^{-1}$, whereas an AGN component is needed in
order to explain the observed X-ray luminosities for the majority of
our sources.  In $\sim$6-13 percent of our sample, we find evidence
for X-ray emission consistent with high SFRs, after accounting for the
relative uncertainties in the L$_{\rm{X}}$-SFR relations and the
typical under-prediction of the 1.1 mm flux in our SED modeling.  The
AGNs typically appear obscured in the X-ray band, with neutral
hydrogen column densities in excess of $10^{23}\rm~cm^{-2}$. These
results are consistent with other SMG/X-ray studies.  Overall, the AGN
templates contribute very little ($\lesssim$10 percent) to both the
bolometric luminosity and 1.1mm flux.  At 1.1 mm in particular, the
AGN+SB models typically under-predict the observed fluxes, which
indicates that either a cooler, extended dust component is required to
fully recover the NIR-to-radio SED or that the sources are blended.
We suggest that this missing dust could result from radiation- and/or
momentum-driven outflows caused by the starburst/AGN regions, which
pushes the dust out into the halo where it cools rapidly and, although
it accounts for a small fraction of the total dust mass, may
contribute significantly to the 1.1 mm emission.  

The high AGN identification rate in these AzTEC SMGs is particularly
interesting in regards to SMG formation and evolution scenarios.
Following a merger-driven scenario, the X-ray identified sources could
represent the transitional period between starburst and AGN dominant
phases.  However, the lack of a significant correlation at large
scales between all X-ray sources and SMGs in these fields suggests
that not all SMGs will evolve to possess an AGN and, similarly, that
not all AGN evolve from a sub-mm bright phase. This suggests
heterogeneity in the formation/evolution of SMGs, possibly due to
either intrinsic source properties, i.e. amount of obscuration, or
even multiple formation scenarios.  With future analyses aimed at
source evolution as a function of redshift, combined with a more
comprehensive redshift catalog for SMGs (one of the goals for the
upcoming LMT), we will be able to determine the AGN fraction and
contribution to greater certainty, allowing for investigating how SMGs
form and evolve into the galaxies we see in the local Universe.

\section*{Acknowledgments}

We would like to thank the AzTEC team, M. Giavalisco, M. Lacy and the
anonymous reviewer for discussions and helpful comments in developing
this manuscript.  This work has been funded under NSF grants
AST-0907952 and AST-0838222, and CXC grant SAO SP1-12003X. M. Kim was
supported in part by Mid-career Researcher Program through the
National Research Foundation of Korea (NRF) funded by the Ministry of
Education, Science and Technology 2011-0028001.  K.S. Scott is
supported by the National Radio Astronomy Observatory, which is a
facility of the National Science Foundation operated under cooperative
agreement by Associated Universities, Inc.

\end{document}